\documentclass{template}

\title{\textbf{Characterization of coherent flow structures in brain ventricles}}
\author[1,2,*]{Halvor Herlyng}
\author[3]{Shawn C.~Shadden}
\affil[1]{Department of Numerical Analysis and Scientific Computing, Simula Research Laboratory, Oslo, Norway}
\affil[2]{Expert Analytics AS, Oslo, Norway}
\affil[3]{Department of Mechanical Engineering, University of California, Berkeley, USA}
\affil[*]{\href{mailto:halvor@xal.no}{halvor@xal.no}}
\date{\vspace{-2em}} 

\begin{document}

\maketitle

\begin{abstract}
    The dynamic flow of cerebrospinal fluid (CSF) in brain ventricles 
    exhibits flow features on several scales, both spatially and temporally.
    Most analysis of this complex flow and the accompanying transport
    has used instantaneous (Eulerian) flow variables. Such analysis
    makes understanding of unsteady transport challenging.
    Here, we analyze brain ventricular CSF flow both in a Eulerian
    sense and from the Lagrangian perspective -- a time-integrated view of the flow.
    With geometries generated from imaging data, we model CSF flow
    in adult human and embryonic zebrafish brain ventricles. 
    In the human brain we model flow governed by cardiovascular pulsations,
    CSF secretion and motile cilia. 
    The flow driven by cardiovascular pulsations is derived from a damped linear elastic model
    of brain ventricle deformations, as a result of applying displacement boundary
    conditions derived from experimental data.
    In the zebrafish brain we consider flow driven solely by motile cilia.
    The tissue and flow models are implemented and solved with finite element methods.
    We use the resulting velocity fields to compute finite-time Lyapunov exponent (FTLE)
    fields and use these fields to characterize Lagrangian coherent structures,
    which can be approximated by ridges in the FTLE fields.
    These coherent structures demonstrate prominent flow features in the brain ventricles
    congruent with findings in experimental research. In the human brain ventricles, we also 
    investigate the role of inertia by comparing flow models governed by 
    the Navier-Stokes and the Stokes equations. Comparisons show that
    solving the Stokes equations is adequate to compute integrated flow
    variables like stroke volumes, but that the Stokes approximation fails 
    to resolve intricate features of flow and advective transport that
    are present in the solution to the Navier-Stokes equations,
    features that could be important to elucidating transport.
\end{abstract}

\section{Introduction}
Brains are constantly in motion. These motions are tightly coupled
to the cardiac and respiratory cycles, bodily movement and
neural activity. This brain motion is a key contributor to
cerebrospinal fluid (CSF) flow~\cite{Enzmann1992BrainImaging,
baledent2001cerebrospinal, sweetman2011cerebrospinal}.
CSF is produced by choroid plexus in the brain ventricles~\cite{Damkier2013},
from where the fluid flows into the spinal cord or
to the subarachnoid space~\cite{rasmussen2018glymphatic}.
The flow of CSF maintains brain homeostasis,
delivers nutrients and removes waste, and
serves as a cushion for the brain~\cite{fame2020emergence,
kelley2023cerebrospinal, linninger2016cerebrospinal}.
Meanwhile, several pathologies afflicting the brain
are related to disruption of CSF flow,
such as hydrocephalus, where ventricular enlargement
results from blockage of CSF flow~\cite{linninger2016cerebrospinal}.
The nature of CSF flow has been studied extensively
in recent decades, both experimentally and computationally.
Among the physiological processes that affect CSF flow are:
blood vessel and respiration dynamics,
secretion and absorption of CSF,
beating of motile cilia, neural activity, and posture changes.
\cite{kelley2023cerebrospinal, sweetman2011cerebrospinal,
Olstad2019CiliaryDevelopment, Faubel2016CiliaVentricles,
Kurtcuoglu2007ComputationalSylvius, Czosnyka2004CerebrospinalDynamics,
Vinje2019RespiratoryMeasurements, Liu2024UsingBreathing,
baledent2001cerebrospinal}.
With these physical processes occurring
over various time and length scales, CSF flow
also consists of motions on different
scales~\cite{kelley2023cerebrospinal}.
Although CSF is fundamental to brain functioning,
the underlying mechanisms of CSF flow are still not
completely understood~\cite{kelley2023cerebrospinal}.

Because of the negative consequences of disruption in
ventricular CSF flow, increasing our understanding of
this flow is important to prevent or treat associated 
pathologies. In the brain ventricles, 
pulsations originating from cardiac and respiratory dynamics
cause perturbations to blood volume and tissue deformation
in the brain that result in 
CSF flow~\cite{enzmann1991normal,
wagshul2011pulsating, Liu2024UsingBreathing, baledent2004relationship,
greitz1992pulsatile, Vinje2019RespiratoryMeasurements,
kelley2023cerebrospinal, greitz1993pulsatile}.
Back-and-forth motion of CSF with vortical flow features
has been visualized in human brains~\cite{stadlbauer2010insight},
and the pulsatility is clearly present in measurements of flow rates
\cite{baledent2004relationship, Liu2024UsingBreathing, Vinje2019RespiratoryMeasurements}.
At smaller scales, visualizations of ventricular
CSF flow trajectories in animals have demonstrated
intricate features such as stagnation and separation points,
and vortex-structured flows created by motile cilia
\cite{gilpin2017flowtrace, Faubel2016CiliaVentricles,
Olstad2019CiliaryDevelopment}.
Cilia-driven CSF flow has also 
to a lesser extent been studied in humans~\cite{worthington1963ependymal,
Yoshida2022EffectVentricles, Siyahhan2014FlowVentricles},
demonstrating local effects of cilia on CSF flow patterns.

Flow patterns such as separatrices and vortices can be elucidated by special
material structures known as Lagrangian coherent
structures (LCS). The LCS constitute transport barriers in
fluid advection, and organize stretching and 
folding in the mixing of passive tracers~\cite{haller2015lagrangian,
haller2000lagrangian}. LCS can be characterized by computing
finite-time Lyapunov exponent (FTLE) fields through integration of fluid
parcel trajectories; ridges of these FTLE fields are good approximations of
LCS~\cite{shadden2005definition}. Characterizing LCS thus provides
a time-integrated view of fluid transport
\cite{shadden2005definition, haller2015lagrangian, haller2000lagrangian}.
The explanatory power of LCS has previously been demonstrated
in a host of applications, including blood flow
\cite{shadden2008characterization, shadden2010computational},
atmospheric flow~\cite{Gunther2021LagrangianWinds},
flow in airways~\cite{lukens2010using}
and oceanic flows~\cite{beron2015dissipative}.

Despite the predictive power of studying flow from a
Lagrangian perspective--where fluid parcel trajectories are tracked over time--only
a limited amount of previous ventricular CSF flow studies
have considered this perspective~\cite{Yoshida2022EffectVentricles}.
Most studies are rather based on flow analysis with an
Eulerian perspective~\cite{Kurtcuoglu2007ComputationalSylvius,
Causemann2022HumanFramework, howden2008three, Sweetman2011ThreeDimensional,
Siyahhan2014FlowVentricles},
in which flow variables such as velocity and volumetric flux
are measured at a fixed point in space and time.
Instantaneous flow variables may however deceive
interpretation of time-dependence in the flow~\cite{shadden2005definition},
especially in flows that contain motion at multiple scales,
such as ventricular CSF flow. Therefore,
further knowledge of flow and transport in brain ventricles,
and the precise contributions of specific
flow mechanisms such as cardiovascular pulsations, CSF secretion,
and beating of motile cilia to mixing and transport of CSF,
could be gained from Lagrangian analysis.
For example, do LCS exist in brain ventricular CSF flow,
and if so, how are the LCS organized?
How are the LCS affected by flow mechanisms
such as cardiovascular pulsations, motile cilia and CSF secretion?
Does fluid inertia play a significant role in establishing the
LCS and thus the advective transport within the brain ventricles?

In this work, we characterize LCS in brain ventricular CSF flow
using two computational models generated with imaging data.
One geometry represents the brain ventricles of an adult human. The other
geometry represents the brain ventricles of embryonic zebrafish, an animal whose
ventricular system shares many similarities with the human brain ventricles.
Using finite element methods, we simulate CSF flow
in these two geometries. The CSF velocity fields are used to
compute particle trajectories. These trajectories are used to
compute finite-time Lyapunov exponent (FTLE) fields, which
we use to characterize LCS. In the human brain, we model
flow driven by cardiovascular pulsations, choroid plexus secretion and motile cilia,
and study the impact of each mechanism on flow patterns and advective transport.
The results indicate the presence of LCS that are mostly determined
by cardiovascular pulsations. For the zebrafish brain, we model flow
driven by motile cilia, and demonstrate that FTLE fields can be
used to characterize LCS that reflect flow trajectories
previously observed in experiments. Altogether, our work demonstrates the usefulness of
Lagrangian analysis in furthering our understanding of CSF flow.

\section{Methods}
\subsection{Image-based computational meshes of ventricular geometries}
\begin{figure}
    \centering
    \includegraphics[width=\textwidth]{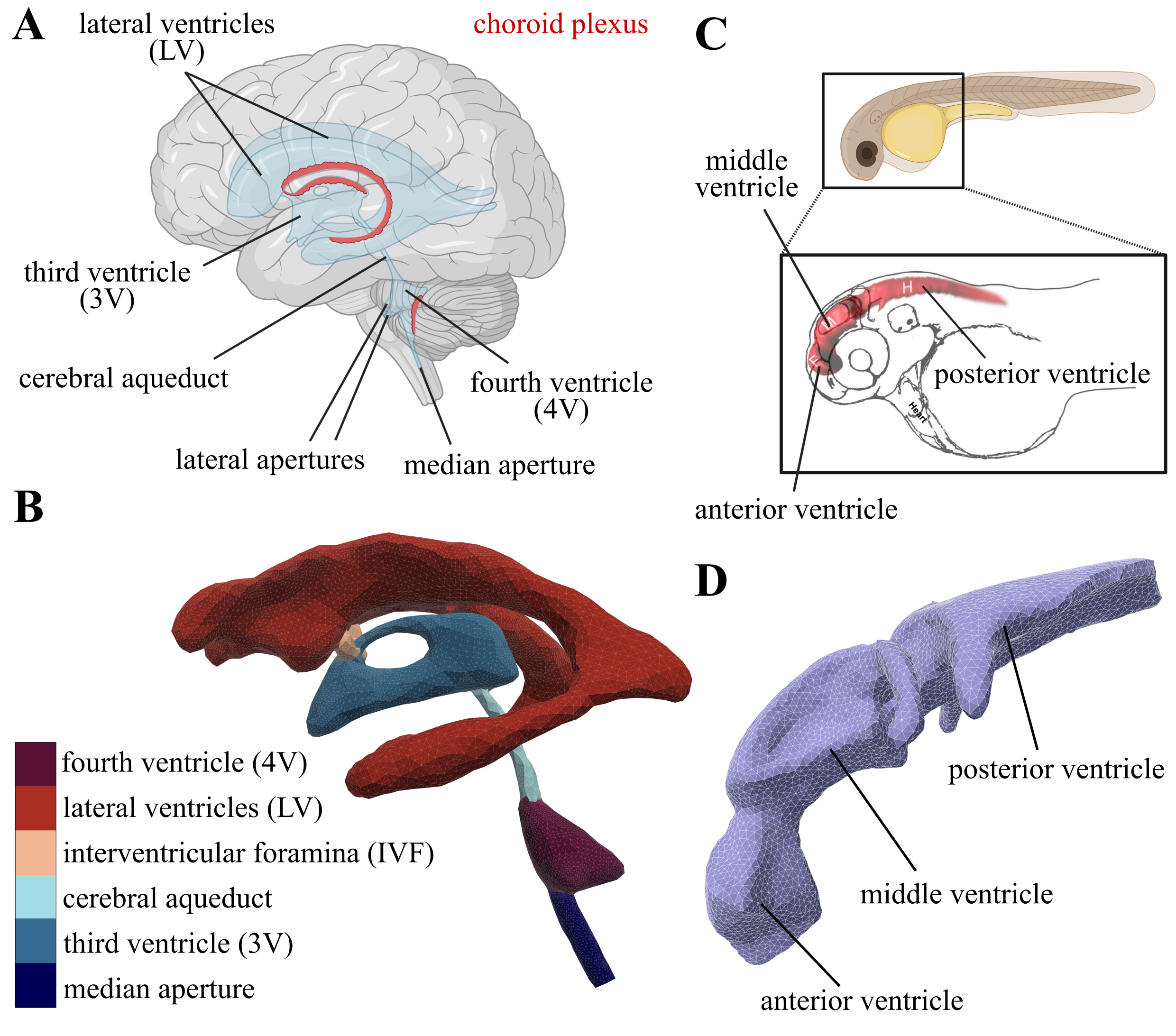}
    \caption{
        \textbf{A.} Illustration of the brain ventricular system
        in the adult human brain, with indication of the
        choroid plexus tissue (red structures) locations
        in the lateral, third and fourth ventricles. 
        \textbf{B.} Computational mesh of human brain ventricles, sagittal view..
        \textbf{C.} Illustration of the brain ventricular system
        in the embryonic zebrafish brain, indicating the location
        of the anterior, middle and posterior ventricles.
        \textbf{D.} Computational mesh of embryonic zebrafish brain ventricles,
        sagittal view.
        Parts of this figure were created with BioRender.
        The sketch of the zebrafish brain in panel C is adapted subject
        to the Creative Commons Attribution 4.0 International License
        (\href{https://creativecommons.org/licenses/by/4.0/}{https://creativecommons.org/licenses/by/4.0/})
        \cite{korzh2018development}.
    }
    \label{fig:model_sketch}
\end{figure}

To study brain ventricular CSF flow and characterize coherent flow structures,
we consider two brain ventricular geometries generated with imaging data
(\Cref{fig:model_sketch}). One is the brain ventricular system of an adult human,
which includes of the lateral, third and fourth ventricles (\Cref{fig:model_sketch}A).
The interventricular foramina connect the lateral ventricles to the third ventricle,
while the cerebral aqueduct connects the third and fourth ventricles (\Cref{fig:model_sketch}A, B).
In the lateral, third and fourth ventricles there are regions covered by choroid plexus tissue
(\Cref{fig:model_sketch}A). The lateral apertures and the median aperture serve as exit routes for the CSF.
To represent the human ventricular system, we use a tetrahedral computational mesh extracted
from the brain mesh originally generated by Causemann~\emph{et al.}~\cite{Causemann2022HumanFramework}.
This brain mesh was originally generated using surface geometries based on MRI images.
Here, the standard version of the human brain ventricles mesh consists of
408 128 tetrahedral cells with maximal and minimal edge lengths
of 0.40 cm and 0.023 cm, respectively.

The second geometry represents embryonic zebrafish brain ventricles
at two days post-fertilization (2 dpf, \Cref{fig:model_sketch}C).
A surface geometry of these ventricles was in previous work generated with confocal
microscopy~\cite{Olstad2019CiliaryDevelopment}. This geometry was used
to generate a tetrahedral computational mesh using fTetWild~\cite{Hu2020FastWild}
(\Cref{fig:model_sketch}D). At 2 dpf, the zebrafish brain
ventricular system consists of three cavities:
the anterior, middle and posterior ventricles (\Cref{fig:model_sketch}C, D).
The three compartments are connected by interventricular ducts. The computational
mesh constitutes 132 134 tetrahedral cells, with a maximal (minimal) edge
length of 14.9 $\mu$m (4.7 $\mu$m).

\subsection{Governing equations of brain ventricular cerebrospinal fluid flow}
We model cerebrospinal fluid (CSF) as an incompressible, Newtonian fluid of density
$\rho=1000 \ \mathrm{kg/m^3}$ and dynamic viscosity
$\mu=0.7 \ \mathrm{mPa\cdot s}$~\cite{Bloomfield1998EffectsFluid}
in a domain $\DD\subset\mathbb{R}^d$
with boundary $\partial\DD\subset\mathbb{R}^{d-1}$.
Neglecting the gravitational acceleration, no body forces are acting on the fluid.
The flow is governed by the Navier-Stokes equations, which determine the
fluid velocity $\uu(\xx, t)$ and pressure $p(\xx, t)$ for times $t > 0$:
\begin{alignat}{2}
    \rho\left(\pdifft{\uu}
    + (\uu\cdot\nabla)\uu\right) &= \nabla\cdot\TT \quad&&\text{in }\DD,
    \label{eq:navier_stokes_mom_eq} \\
    \nabla\cdot\uu &= 0 \quad&&\text{in }\DD,
    \label{eq:continuity_eq}
\end{alignat}
with the stress tensor $\TT$ defined as
\begin{equation*}
    \TT = 2\mu\beps(\uu) - p\mathbb{I}.
\end{equation*}
Here $\mathbb{I}$ is the identity tensor of dimension $d$ and
$\beps(\uu)=\frac{1}{2}\left(\nabla\uu + (\nabla\uu)^T\right)$ is the
strain-rate tensor.

For fluid flows where viscous forces dominate inertial 
forces, the steady Stokes equations, which read
\begin{equation}
    - \nabla\cdot\TT = \bm{0} \quad\text{in }\DD,
    \label{eq:stokes_eq}
\end{equation}
are often used as an approximation to the momentum equation
(\Cref{eq:navier_stokes_mom_eq}).
Viscous forces dominate in CSF flow in smaller brains
such as embryonic zebrafish brain ventricles
\cite{Olstad2019CiliaryDevelopment, Herlyng2025AdvectionDiffusion},
and therefore we use \Cref{eq:stokes_eq} to model CSF flow
in zebrafish. In the human brain, flow measurements and simulations
indicate that CSF flow is in a laminar regime
with non-negligible transient effects, but negligible
inertial effects~\cite{Gholampour2018FSI, howden2008three}.
In such a flow regime, the unsteady Stokes equations
\begin{equation}
    \rho\pdifft{\uu} - \nabla\cdot\TT = \bm{0} \quad\text{in }\DD,
    \label{eq:unsteady_stokes_eq}
\end{equation}
may serve as a fair approximation.
In~\Cref{eq:unsteady_stokes_eq}, the convective acceleration term
$\rho (\uu\cdot\nabla)\uu$ in~\Cref{eq:navier_stokes_mom_eq} is neglected,
while the local acceleration
$\rho\partial\uu/\partial t$ is retained to account for transient effects.
Although most computational studies of CSF flow in human brain ventricles
employ the Navier-Stokes momentum equation, some previous
studies are based on \Cref{eq:unsteady_stokes_eq}
\cite{Causemann2022HumanFramework, kelley2023cerebrospinal}.
We will compute flow in human brain ventricles with both the
Navier-Stokes (\Cref{eq:navier_stokes_mom_eq}) and the unsteady
Stokes (\Cref{eq:unsteady_stokes_eq}) momentum equations to assess
the role of fluid inertia, and whether the unsteady Stokes equations
serve as a fair approximation in this setting.

\subsection{Governing equations of brain tissue deformation}
For the human brain ventricles, we use displacements of the brain
ventricles walls as a driving force for CSF flow. The wall displacements are 
derived from a model of brain tissue deformation that considers the brain
ventricles as a linear elastic material with viscoelastic
behavior modeled by Rayleigh damping. The displacement $\ww(\xx, t)$
of the brain ventricles is then governed by the linear elastodynamics
equations
\begin{equation}
    \rho\ddot{\ww} + \zeta_M\rho\dot{\ww} - \zeta_K\nabla\cdot\TT_t(\dot{\ww})
    - \nabla\cdot\TT_t(\ww) = \bm{0} \quad\text{in }\DD,
    \label{eq:strong_linear_elasticity_1}
\end{equation}
where $\dot{\ww} = \pdifft{\ww}$ and $\ddot{\ww} = \pdiffdifft{\ww}$
are the displacement velocity and acceleration, respectively.
The two damping factors $\zeta_M$ and $\zeta_K$
are related to inertia and stiffness. 
Moreover, $\TT_t = 2\eta\beps(\ww) + \lambda\mathrm{tr}(\beps(\ww))\delta$ is the
tissue stress tensor, in which $\eta$ and $\lambda$ are the Lamé parameters,
and $\delta$ is the Kronecker delta. 

The Lamé parameters $\eta$ and $\lambda$ are derived from
Poisson's ratio $\nu$ and Young's modulus $E$ of brain tissue.
In~\Cref{eq:strong_linear_elasticity_1}, the domain $\DD$
corresponds to the fluid domain, so that a deformation field
can be computed over this region and used to smoothly
adjust the volumetric fluid mesh without large distortion
of mesh elements near the boundary. 
Since this region is not tissue per se,
we expect the precise choice of $\nu$ and $E$ not to
influence the general trends of the results, but rather
only the required magnitude of the applied displacements $\bm{g}(\xx, t)$.
Additionally, there is in general an uncertainty
in these material parameters. Especially $E$, 
for which values reported in the literature span several orders of magnitude
\cite{Budday2015Mechanical, Cheng2010ComputationalModel,
Cheng2007UnconfinedCompression, Sweetman2011ThreeDimensional}.
We adopt the choices of $\nu=0.479$ and $E=1500$
made by Causemann~\emph{et al.}~\cite{Causemann2022HumanFramework}
in their computational model of pulsatile CSF flow in the brain. The $\nu$ value 
reflects a near incompressibility of brain tissue and the ventricular system,
while the $E$ value represents an average of
gray and white matter brain tissue surrounding the ventricular system
\cite{Budday2015Mechanical}. 

In general, the total Rayleigh damping factor $\zeta$ depends on
two separate damping factors, one related to inertia ($\zeta_M$)
and one related to stiffness ($\zeta_K$), and is given by
$\zeta = \zeta_M / (2\omega) + \zeta_K\omega / 2$
for some eigenfrequency $\omega$~\cite{ChopraAnilK2007Dynamics}.
The mass and stiffness damping factors
$\eta_M = 0.20$ and $\eta_K = 0.10$ were set to dampen the motion
such as to control vibrations around the cardiac cycle frequency 1 Hz,
which is close to some of the eiqenfrequencies of the geometry.

\subsection{Modeling CSF flow in deforming brain ventricles}
The tissue displacements modeled by \Cref{eq:strong_linear_elasticity_1}
are used as a driving force for CSF flow by applying displacement velocities
$\dot{\ww}(\xx, t)$ as boundary conditions in the fluid model.
To account for the fact that the fluid then occupies a deforming domain,
we solve the Navier-Stokes equations (\Cref{eq:navier_stokes_mom_eq,eq:continuity_eq})
and the unsteady Stokes equations (\Cref{eq:continuity_eq,eq:unsteady_stokes_eq})
with an Arbitrary Lagrangian Eulerian (ALE)
method~\cite{Duarte2004ALE,Scovazzi2007LectureNotes}.
The governing equations of CSF flow are solved in the reference configuration
$\DD_{\chi} = \DD(\xx, t_0)$, using an invertible mapping
$\psi : \DD_{\chi}\mapsto\DD_{\xx}$ between the reference configuration
and the current (deformed) configuration $\DD_{\xx} = \DD(\xx, t)$. 
The mapping $\psi$ induces a transformation of the differential operators
in the governing fluid equations. Similarly, integrals of the weak formulation
must be transformed appropriately to account for the
domain motion~\cite{Scovazzi2007LectureNotes}.

\subsection{Flow mechanisms and boundary conditions}
CSF flow in adult human brain ventricles differs from CSF flow in embryonic
zebrafish brain ventricles. Therefore, we consider different flow mechanisms
and boundary conditions for the two applications.
In the human brain, we consider three flow mechanisms:
\begin{itemize}
    \item Deformation of the brain ventricles related to cardiovascular pulsatility,
    modeled by imposing a displacement velocity
    on the ventricular walls (see~\Cref{subsubsec:deformation})
    \item Secretion of CSF from choroid plexus tissue,
    modeled by imposing a velocity normal to the wall
    which reflects a total influx of CSF (see~\Cref{subsubsec:secretion})
    \item Ciliary beating, modeled by a tangential traction
    acting on the fluid at the boundary (see~\Cref{subsubsec:cilia})
\end{itemize}

In the embryonic zebrafish brain ventricles, we consider ciliary beating as
the only flow mechanism mainly for two reasons. First, an extensive description
of cardiovascular pulsatility, similar to the one for human brain ventricles,
does not exist~\cite{Herlyng2025AdvectionDiffusion}.
Second, in zebrafish choroid plexus secretion begins later than 
the developmental stage we consider~\cite{bill2008development}. 

The following three subsections detail our modeling approach for the three
types of flow mechanisms outlined above. For a complete specification of
the resulting CSF flow boundary conditions, see~\Cref{sec:boundary_conditions_human} (human model)
and~\Cref{sec:boundary_conditions_zfish} (zebrafish model).

\subsubsection{Modeling cardiovascular pulsation-related deformation of brain ventricles}
\label{subsubsec:deformation}
The main driver of intraventricular CSF flow in humans is pulsatile volume displacement
of the ventricles, which is closely related to the cardiac and respiratory cycles~\cite{wagshul2011pulsating,
Vinje2019RespiratoryMeasurements, sweetman2011cerebrospinal, baledent2001cerebrospinal, Neumaier2025}. 
During systole, the blood vessels in the brain expand and concomitantly
the parenchyma, putting pressure on the ventricular system. The inward displacement
of the brain ventricles leads to antegrade (from cephalic to caudal) CSF flow.
When the blood vessels relax and contract during diastole, the flow reverses and
CSF flows back into the system~\cite{sweetman2011cerebrospinal, greitz1992pulsatile}.
The respiratory cycle also influences CSF flow pressure 
gradients, but at different frequencies
than the cardiac cycle~\cite{Vinje2019RespiratoryMeasurements}.
In this study, the respiratory influence on brain deformations is not considered,
and we model deformations of the ventricles with a cardiac cycle frequency.

We model CSF flow driven by cardiovascular pulsations
through imposing a velocity boundary condition on the
brain ventricular walls. This velocity boundary condition
is given by the displacement velocities $\dot{\ww}$ obtained 
by solving~\Cref{eq:strong_linear_elasticity_1}.
In our model, the displacement velocity determines the normal component
of the CSF velocity
\begin{equation}
    \uu\cdot\nn = \dot{\ww}\cdot\nn \quad\text{on }\Scal_1,
\end{equation}
where $\nn$ is the outward facing normal and 
$\Scal_1\subset\Scal = \partial\DD$.

\begin{figure}
    \centering
    \includegraphics[width=\textwidth]{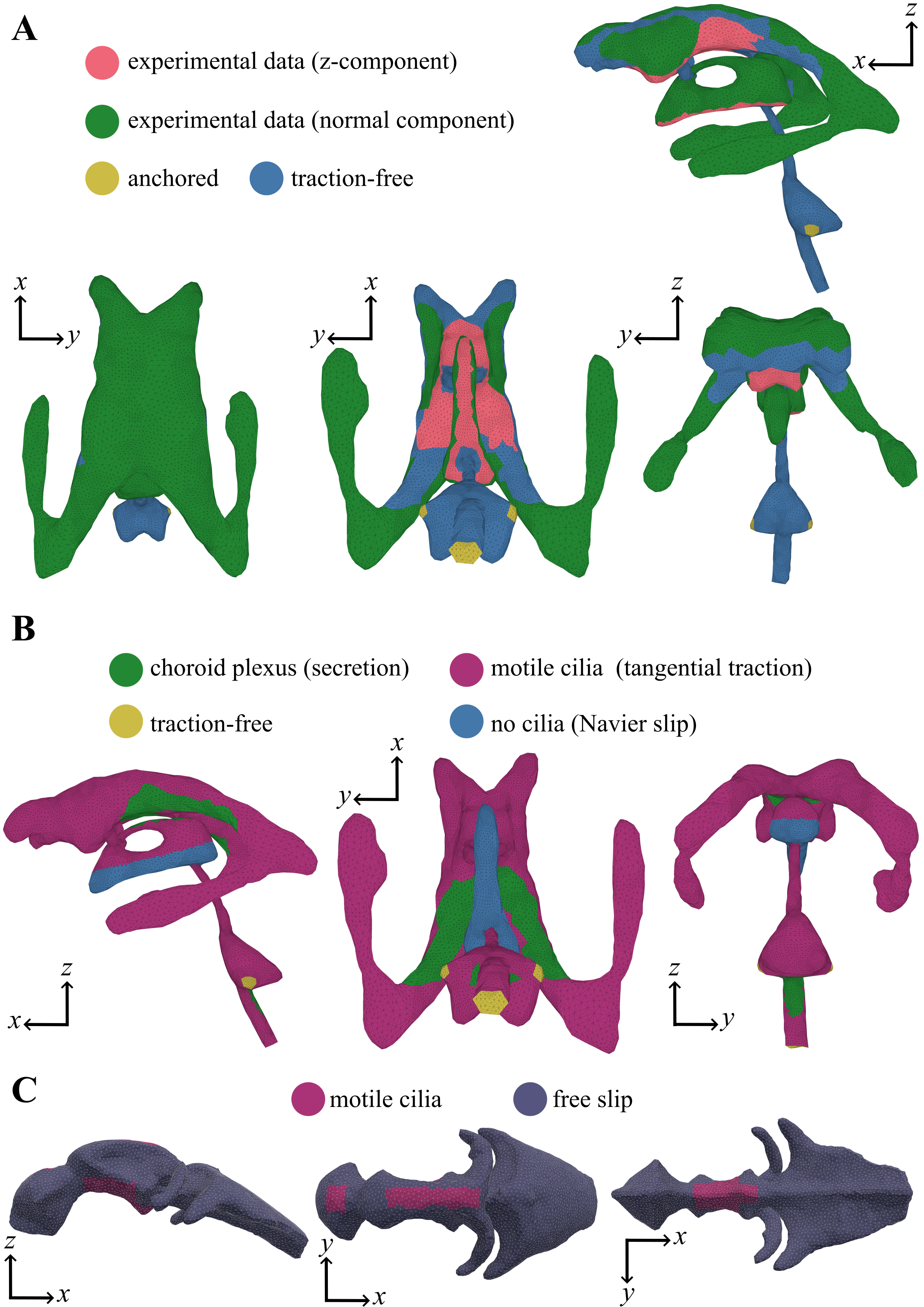}
    \caption{\textbf{A.} Human brain ventricles mesh with boundary
    facets colored by boundary condition type for the tissue deformation
    model.
    \textbf{B.} Human brain ventricles mesh with
    boundary facets colored by boundary condition type for the 
    CSF flow model.
    \textbf{C.} Embryonic zebrafish brain ventricles mesh
    with boundary facets colored by boundary condition
    type for the CSF flow model.}
    \label{fig:mesh_facet_tags}
\end{figure}

Displacement boundary conditions based on experimental
data are used to solve \Cref{eq:strong_linear_elasticity_1} 
and determine $\dot{\ww}$. To this end, we partition the boundary 
$\partial\DD$ into four parts
$\partial\DD=\Bcal_1\cup\Bcal_2\cup\Bcal_3\cup\Bcal_4$ (\Cref{fig:mesh_facet_tags}A)
and impose the boundary conditions
\begin{alignat}{3}
    \ww\cdot\nn  &= \bm{g}\cdot\nn, &&\quad\TT_{t, \parallel} = \bm{0} &&\quad\text{on }\Bcal_1, \label{eq:experimental_normal_displacement_bc}\\
    w_z  &= g_z, &&\quad\TT_t\nn\cdot \bm{e}_x = \TT_t\nn\cdot \bm{e}_y = 0 &&\quad\text{on }\Bcal_2, \label{eq:experimental_z_displacement_bc}\\
    \ww  &= \bm{0} && &&\quad\text{on }\Bcal_3, \label{eq:anchored_displacement_bc}\\
    \TT_t\nn &= \bm{0} && &&\quad\text{on }\Bcal_4, \label{eq:traction_free_displacement_bc}
\end{alignat}
where $\bm{g} := \bm{g}(\xx, t) = (g_x, g_y, g_z)$ is determined from experimental
data and $\bm{e}_i$ is the unit basis vector along the axis $i$.
We have also here introduced the tangential traction vector
$\TT_{\parallel} = P_{\nn}(\TT\nn)$, where
 $$P_{\nn}(\bm{q}) = (\mathbb{I} - \nn\otimes\nn)\bm{q},$$
is the projection of a vector $\bm{q}$ onto a plane normal
to the boundary normal vector $\nn\coloneqq\nn(\xx, t)$.
This projection thus yields a vector tangential to the boundary.

The boundaries $\Bcal_1\cup\Bcal_2$ and imposed data consist of:
\begin{enumerate}[(i)]
    \item The roof of the lateral ventricles, where corpus callosum displacement data is imposed;
    \item The lateral sides of the frontal horns, where caudate nucleus displacement data is imposed;
    \item The occipital and posterior horns, where we impose a systolic contraction followed by a
    diastolic expansion;
    \item The walls of the third ventricle, where thalamus displacement data is imposed;
    \item The floors of the lateral and third ventricles, where we impose cephalocaudal motion.
\end{enumerate}
For (i)--(iv) we impose motion normal to the boundary (\Cref{eq:experimental_normal_displacement_bc},
green facets in \Cref{fig:mesh_facet_tags}A),
while for (v) only motion in the $z$ direction---which aligns with the feet-head direction---
is imposed (\Cref{eq:experimental_z_displacement_bc},
pink facets in \Cref{fig:mesh_facet_tags}A).

The function $\bm{g}(\xx, t)$ was determined by experimental
data~\cite{Kurtcuoglu2007ComputationalSylvius,
Enzmann1992BrainImaging, soellinger20093d, soellinger2007assessment,
greitz1992pulsatile, almudayni2023magnetic, pahlavian2018regional},
and we remark that the function is periodic in time with a period of
one second (assumed to be one cardiac cycle).
For some of the experimental data, directly applying the data produced
deformations that resulted in physically unrealistic large amounts of CSF flow.
Therefore, we reduced the magnitudes of some of the displacements,
while retaining the qualitative behavior
of a downward and contractile motion during systole and an upward
and dilated motion during diastole. Namely, reduction of the magnitudes was 
calibrated by comparing the CSF flow results with experimental values
of aqueductal stroke volume\cite{eide2021direction, wagshul2011pulsating}. 

The boundary $\Bcal_3$ consists of the median aperture outlet
and the lateral apertures (yellow facets in \Cref{fig:mesh_facet_tags}A).
These boundaries are anchored for two reasons:
First, to avoid rigid body motions when solving~\Cref{eq:strong_linear_elasticity_1};
Second, because these boundaries are
considered open boundaries in the fluid model, the displacement velocity
$\dot{\ww}$ on this boundary must be zero to avoid conflicting boundary conditions
between the tissue and the fluid models.
Finally, to let the remaining part of the boundary deform freely subject to the boundary conditions
on $\Bcal_1\cup\Bcal_2\cup\Bcal_3$, the boundary $\Bcal_4$ is traction-free in all directions
(blue facets in \Cref{fig:mesh_facet_tags}A).

\subsubsection{Modeling choroid plexus CSF secretion with a normal flux}\label{subsubsec:secretion}
In the human brain, CSF is mainly produced by secretion
through the choroid plexus tissue in the lateral,
third and fourth ventricles (\Cref{fig:model_sketch}A, B)~\cite{Damkier2013}.
We model CSF secretion by imposing a volumetric flux normal to the
choroid plexus boundary, which is denoted $\Scal_2\subset\Scal$
(green facets in \Cref{fig:mesh_facet_tags}B).
We assume that all CSF that enters the brain ventricles
is produced by the choroid plexus, and that the influx is evenly distributed
over the secretion sites in the lateral, third and fourth ventricles.
A total daily CSF production of $Q_{\mathrm{chp}} = 504$ ml/day is assumed
\cite{Czosnyka2004CerebrospinalDynamics, huff2017neuroanatomy}.
Since we assume even distribution at the different
choroid plexuses, we impose 
\begin{equation}
    \uu\cdot\nn = -q_{\mathrm{chp}} + \dot{\ww}\cdot\nn \quad\text{on }\Scal_2,
    \label{eq:choroid_plexus_BC}
\end{equation}
where $q_{\mathrm{chp}} = Q_{\mathrm{chp}}/S$ and
$S$ is the total choroid plexus boundary surface area.
The normal displacement velocity
$\dot{\ww}\cdot\nn$ is included to account
for the deforming ventricular walls.

\subsubsection{Modeling ciliary beating in the human brain by tangential traction}\label{subsubsec:cilia}
Motile cilia are hair-like structures that line
the ependyma of brain ventricular walls, and the beating motion of cilia propel
CSF~\cite{Ringers2020Role, Olstad2019CiliaryDevelopment}.
Motile cilia are abundant in the human brain~\cite{doetsch1997cellular,
scott1973scanning, coletti2018characterization, worthington1963ependymal},
but are not found in the caudal third of the third ventricle~\cite{scott1972comparative,
bruni1972scanning, sanchez2024exploring} nor on choroid plexus~\cite{scott1972comparative,
ho2023choroid, narita2015cilia}, although these latter areas
may exhibit immotile cilia. 
Based on experimental research, we model cilia everywhere in the 
human brain ventricles, except for in the caudal third of the third
ventricle and on choroid plexus~\cite{doetsch1997cellular,
scott1973scanning, coletti2018characterization, worthington1963ependymal,
scott1972comparative, bruni1972scanning, sanchez2024exploring,
scott1972comparative, ho2023choroid, narita2015cilia}.

The cilia are modeled by imposing on the fluid a
tangential traction $\btau(\xx, t) = \tau P_{\nn}(\rr(\xx, t))$
that represents the collective beating motion of cilia,
where $\tau$ is a constant that determines the 
magnitude of the tangential traction $\btau$.
In the human model, the tangential traction is
imposed through a Navier slip boundary condition:
\begin{equation}
    \TT_{\parallel} = \btau(\xx, t) - \alpha P_{\nn}(\uu)
    \quad\text{on }\Scal_1\setminus\Scal_4,
    \label{eq:cilia_navier_slip_bc}
\end{equation}
where the boundary $\Scal_4\subset\Scal_1$ is the lower third of
the third ventricle (blue facets in \Cref{fig:mesh_facet_tags}B),
where we do not impose cilia, and
$\TT_{\parallel} = P_{\nn}(\TT\nn)$ is the
tangential traction vector. The friction coefficient $\alpha=\mu/L$ 
is determined by the dynamic viscosity $\mu$ and a slip length $L$.
We model the cilia layer lining the ventricular walls
as a slip layer, such that the fluid slips at the cilia while
approaching a no-slip boundary condition at the underlying
ventricular wall. We choose $L=15$ $\mu$m
which is a typical length of a cilium in human brain ventricles
\cite{Siyahhan2014FlowVentricles,Afzelius2004CiliaRelatedDiseases}
as the slip length, resulting in $\alpha=47 \ \mathrm{Pa\cdot s/m}$.
In the regions where cilia are absent, we consider
\Cref{eq:cilia_navier_slip_bc} with $\btau(\xx, t) = \bm{0}$:
\begin{equation}
    \TT_{\parallel} = - \alpha P_{\nn}(\uu)
    \quad\text{on }\Scal_2\cup\Scal_4.
    \label{eq:navier_slip_bc}
\end{equation}
Thus, although we assume motile cilia are absent on $\Scal_2\cup\Scal_4$,
we consider the ventricular walls of these regions to impose friction on the CSF.
The regions $\Scal_2$ and $\Scal_4$ are illustrated by
green and blue facets in \Cref{fig:mesh_facet_tags}B, respectively.

For the human model, we model the direction $\rr$
motivated by the work of Guirao~\emph{et al.},
who found the motile-cilia beating direction to be oriented by the
CSF production during neural development~\cite{Guirao2010CouplingCilia}.
To generate the vector $\rr(\xx, t)$, we simulate CSF flow in a production-only
setup, with the choroid plexus boundary condition~\Cref{eq:choroid_plexus_BC},
a zero normal traction condition on the median and lateral apertures,
and $\uu\cdot\nn=0$ on the rest of the boundary (that is, without wall displacements). 
The resulting velocity field,
$\uu_{\mathrm{prod}}$, is normalized by its Euclidean length
to yield the direction of the cilia:
\begin{equation*}
    \rr(\xx, t) = \frac{\uu_{\mathrm{prod}}(\xx, t)}{|\uu_{\mathrm{prod}}(\xx, t)|}.
\end{equation*}

There does not exist accurate data for the parameter $\tau$ that
determines the magnitude of the tangential traction $\btau$, with only a few computational
studies reporting a value for a cilia force magnitude
\cite{Herlyng2025AdvectionDiffusion, Siyahhan2014FlowVentricles,
Thouvenin2020OriginCanal}. None of these report values for humans.
We set $\tau=7.9$ mPa based on the force magnitude used by
Siyahhan~\emph{et al.}~\cite{Siyahhan2014FlowVentricles}
for the purpose of comparison
since they also modeled CSF flow in human brain ventricles,
despite this value of $\tau$ being based on rodent brain data.

\subsubsection{Summary of the boundary conditions
for the governing CSF flow equations in human brain ventricles}
\label{sec:boundary_conditions_human}

For the human CSF flow model, the boundary is partitioned into
four parts: $\partial\DD = \Scal = \Scal_1\cup\Scal_2\cup\Scal_3\cup\Scal_4$.
Using the mechanisms introduced in the previous subsections,
the boundary conditions are
\begin{alignat}{2}
    \uu\cdot\nn &= \dot{\ww}\cdot\nn &&\quad\mathrm{on \ \mathcal{S}_1}, \\
    \TT_{\parallel} &= \tau P_{\nn}(\rr) - \alpha P_{\nn}(\uu)
    &&\quad\mathrm{on \ \mathcal{S}_1\setminus\mathcal{S}_4}, \label{eq:cilia_bc} \\
    \uu\cdot\nn &= -q_{\mathrm{chp}} + \dot{\ww}\cdot\nn &&\quad\mathrm{on \ \mathcal{S}_2}, \\
    \TT_{\parallel} &= - \alpha P_{\nn}(\uu) &&\quad\mathrm{on \ \mathcal{S}_2\cup\mathcal{S}_4}, \label{eq:navier_bc} \\
    \TT\nn &= \bm{0} &&\quad\mathrm{on \ \mathcal{S}_3}. \label{eq:outlet_bc}
\end{alignat}
We remark that the boundary $\mathcal{S}_4\subset\mathcal{S}_1$ is the lower third of
the third ventricle, where we do not impose the cilia tangential traction $\btau$
as we consider motile cilia to be absent in these regions based on experimental
observations. Moreover, $\dot{\ww}(\xx, t)$ is the displacement velocity of the ventricular walls,
the vector $\TT_{\parallel}=P_{\nn}(\TT\nn)=\TT\nn - (\TT\nn\cdot\nn)\nn$
is the tangential traction, and $P_{\nn}(\cdot)$ is the projection
of a vector onto a plane normal
to the outward-facing normal vector $\nn$. 
\Cref{eq:outlet_bc} imposes zero traction,
and is imposed on the end of the median aperture and on the lateral apertures
to simulate open boundaries.

\subsubsection{Boundary conditions for CSF flow in embryonic zebrafish brain ventricles}
\label{sec:boundary_conditions_zfish}
For the zebrafish model, the only flow mechanism
we consider is motile cilia. We let the domain boundary be denoted
$\mathcal{Z} = \mathcal{Z}_1\cup\mathcal{Z}_2$.
Similar to previous work~\cite{Herlyng2025AdvectionDiffusion},
we consider a frictionless ($\alpha=0$) 
tangential traction on the cilia boundary
and a free-slip condition elsewhere:
\begin{alignat}{2}
    \TT_{\parallel} &= \Tilde{\btau}(\xx, t)  &&\quad\text{on }\mathcal{Z}_1, \\
    \TT_{\parallel} &= \bm{0} &&\quad\text{on }\mathcal{Z}_2, \\
    \uu\cdot\nn &= 0 &&\quad\text{on }\mathcal{Z}.\label{eq:impermeability_bc_zfish}
\end{alignat}
Condition \eqref{eq:impermeability_bc_zfish} reflects the assumption that the
ventricular walls are impermeable. The cilia
boundary $\mathcal{Z}_1$ covers parts of the dorsal and ventral
regions of the middle ventricle walls, as well as the dorsal wall of the anterior
ventricle~\cite{Olstad2019CiliaryDevelopment} (purple facets in \Cref{fig:mesh_facet_tags}C).
The function $\Tilde{\btau}(\xx, t)$ yields stronger cilia forces in the middle ventricle
and weaker forces in the anterior ventricle, and was determined based on
experimental data~\cite{Olstad2019CiliaryDevelopment}. We refer the reader to
Herlyng~\emph{et al.}~\cite{Herlyng2025AdvectionDiffusion} for more details.

\subsection{Finite-time Lyapunov exponent fields and Lagrangian coherent structures}
We computed finite-time Lyapunov exponent (FTLE) fields to
identify Lagrangian coherent structures (LCS).
LCS are transport barriers that partition regions of distinct flow behavior
and act as organizing structures of fluid advection.
LCS can be identified as ridges in the FTLE field~\cite{shadden2005definition}. 

FTLE fields can be computed from fluid particle trajectory information~\cite{shadden2005definition}.
Fluid particle trajectories $\xx(t)\in\mathcal{D}\subset\mathbb{R}^d$
can be described by the dynamical system
\begin{equation}
    \dot{\xx}(t; t_0, \xx_0) := \frac{\mathrm d\xx}{\mathrm d t} = \uu(\xx(t; t_0, \xx_0), t),
    \label{eq:dx_dt_equals_u}
\end{equation}
where $t$ denotes time, $\uu$ is the CSF velocity, and $\xx_0=\xx(t_0; t_0, \xx_0)$
is the initial position of a fluid particle at the initial time $t_0$. 
\Cref{eq:dx_dt_equals_u} can be integrated to yield the particle position at a time
$t$:
\begin{equation}
    \xx(t; t_0, \xx_0) = \xx(t_0) + \int_{t_0}^t \uu(\xx, \xi)\,\mathrm d\xi.
    \label{eq:integrate_dx_dt}
\end{equation}
This relation can be used to define the flow map
\begin{equation}
    \bphi_{t_0}^t : \mathcal{D} \rightarrow \mathcal{D} : \xx_0 \mapsto \bphi_{t_0}^t(\xx_0) = \xx(t; t_0, \xx_0).
\end{equation}
Using the gradient of the flow map,
we define the (right) Cauchy-Green deformation tensor
\begin{equation}
    \mathbf{C} = \left(\nabla\bphi_{t_0}^t\right)^*\nabla\bphi_{t_0}^t,
\end{equation}
where $\mathbf{A}^*$ denotes the transpose of a tensor $\mathbf{A}$.
The finite-time Lyapunov exponent (FTLE) is defined as
\begin{equation}
    \sigma:=\sigma_{t_0}^{t_0+T}(\xx_0) = \frac{1}{\vert T\vert}\ln{\sqrt{\lambda_{\mathrm{max}}}},
\end{equation}
where we have defined the integration time $T = t - t_0$, and $\lambda_{\mathrm{max}}$
is the maximum eigenvalue of $\mathbf{C}$~\cite{shadden2005definition}.
The FTLE measures how sensitive a fluid trajectory
is to perturbations in its initial position.
Namely, let $\boldsymbol{\delta}(t_0)$ be an infinitesimal change
in position from a fluid particle at $\xx_0$ at time $t_0$.
This perturbation will have at most an exponential growth
rate of $\sigma$ according to~\cite{shadden2008characterization}
\begin{equation}
    \max||\boldsymbol{\delta}(t_0 + T)||
    \leq e^{\sigma |T|}||\boldsymbol{\delta}(t_0)||.
\end{equation}
Note that the absolute value of the integration time $T$
is employed because FTLE values can be computed
both by integration forward ($t>t_0$) and backward ($t<t_0$) in time.
FTLE values computed by integration forward in time
reveal repelling LCS of the flow,
whereas FTLE values computed by integration backward in time 
reveal attracting LCS. 

\subsection{Model scenarios and quantities of interest}
To quantify flow and analyze the role of different flow mechanisms
in the human brain model, we investigate three model scenarios:
DSC includes all three flow mechanisms (deformation, secretion, cilia);
DC includes deformation and cilia, excludes secretion;
DS includes deformation and secretion, excludes cilia.
To compare results for the different mechanisms, and
also for comparison with previous computational and clinical studies,
we compute the following quantities:
\begin{itemize}
    \item The volumetric CSF flow rate
    $$ Q(t) = \int_{\mathcal{C}}\uu(\xx, t)\cdot\nn(\xx)\,\mathrm dS $$
    over different cross sections $\mathcal{C}$ in the interventricular foramina and cerebral aqueduct, 
    as well as the outlets of the lateral apertures and median aperture.

    \item The cumulative CSF flow volume through the same cross sections
    $\mathcal{C}$ in the aqueduct,
    median aperture, lateral apertures
    and interventricular foramina. The cumulative flow volume $V(t)$
    is computed by integrating the volumetric flow rate $Q(t)$
    from an initial time $t_0$ to time $t$:
    $$ V(t) = \int_{t_0}^{t} Q(\xi)\,\mathrm d\xi.$$

    \item The CSF stroke volume in the aqueduct, median aperture, lateral apertures,
    and interventricular foramina. The stroke volume is the maximum (with respect
    to time) of the cumulative flow volume.

    \item The peak-to-nadir mean pressure gradient in the cerebral aqueduct,
    measured as the difference in the maximum and minimum value of
    the mean pressure gradient $\Delta \overline{p}$ over a cardiac cycle.
    Here, the mean denotes the cross-sectional average of the pressure:
    $$\overline{p}(t) = \frac{1}{|\mathcal{C}|}\int_{\mathcal{C}} p(\xx, t)\,\mathrm dS,$$
    where $\mathcal{C}$ is a given cross section with area $|\mathcal{C}|$.
    Moreover, the $\Delta$ signifies the change in the mean pressure over the
    aqueduct length:
    $$ \Delta \overline{p} = \frac{\overline{p}_{\mathrm{top}} - \overline{p}_{\mathrm{bottom}}}{L_{\mathrm{aq}}},$$
    where ``top'' and ``bottom'' denote cross sections at the top and bottom of the aqueduct,
    while $L_{\mathrm{aq}} = 23.5$ mm is the length that separates these cross sections.

    \item Maximum ($u_{\mathrm{max}}$) and minimum ($u_{\mathrm{min}}$)
    CSF velocity magnitudes over one cardiac cycle.
    At a given time $t$, the magnitude $u$ of the velocity vector
    $\uu(\xx, t) = (u_x(\xx, t), u_y(\xx, t), u_z(\xx, t))$ is 
    $$ u(\xx, t) = \sqrt{u_{x}^2 + u_{y}^2 + u_{z}^2}. $$
\end{itemize}

\subsection{Numerical methods}
\label{sec:numerical_methods}

This section provides details on the numerical methods used
to solve the governing equations for brain tissue deformation and CSF flow,
as well as the computation of FTLE fields.
To summarize, the general algorithm of the computational model of the human
brain ventricles is:
\begin{enumerate}
    \item Solve the elasticity equations for the brain tissue displacements $\ww(\xx, t)$
    and use $\ww$ to compute the wall displacement velocity $\dot{\ww}(\xx, t)$,
    which is used as a boundary condition in the CSF flow model
    \item Solve the fluid equations to obtain the CSF velocity field $\uu(\xx, t)$ and pressure $p(\xx, t)$
    \item Use the velocity $\uu(\xx, t)$ and pressure $p(\xx, t)$ to compute quantities of interest
    \item Use the velocity $\uu(\xx, t)$ to compute FTLE fields
\end{enumerate}
The algorithm of the zebrafish model consists of steps 2 to 4.

\subsubsection{Finite element method for the tissue deformation equations}
The unsteady (damped) linear elasticity equations (\Cref{eq:strong_linear_elasticity_1})
that govern the brain tissue deformation are discretized in space with a finite element method 
using fourth-order continuous Lagrange elements. For the displacement boundary conditions,
we imposed traction-free conditions (\Cref{eq:traction_free_displacement_bc})
weakly. For the boundary conditions given by displacements $\bm{g}(\xx, t)$ based on experimental data,
the feet-head displacements \eqref{eq:experimental_z_displacement_bc} were imposed strongly,
while the normal displacements \eqref{eq:experimental_normal_displacement_bc}
were imposed weakly using Nitsche's method~\cite{Nitsche1971UberSind,
benzaken2024constructing}. The anchor conditions \eqref{eq:anchored_displacement_bc}
were imposed strongly.

In time, the tissue deformation equations are discretized with the Newmark-$\beta$ method,
for which updates of the displacement $\ww$ and 
displacement velocity $\dot{\ww}$ are
\begin{equation}
    \begin{aligned}
        \ww_{n+1} &= \ww_n + \Delta t\dot{\ww}_n + \left(\frac{1}{2} - \beta\right)\Delta t^2\ddot{\ww}_n + \beta\Delta t^2\ddot{\ww}_{n+1}, \\
        \dot{\ww}_{n+1} &= \dot{\ww}_n + \left(1 - \gamma\right)\Delta t\ddot{\ww}_n + \gamma\Delta t\ddot{\ww}_{n+1},
    \end{aligned}
    \label{eq:newmark_beta_method}
\end{equation}
where $n$ denotes time step and $\Delta t = 0.001$ is the time step size,
while $\ddot{\ww}_n$ is the acceleration.
We set $\gamma=1/2$ and $\beta=1/4$, which yields
a second-order accurate and unconditionally
stable scheme~\cite{newmark1959method}.
With this choice of $\gamma$ and $\beta$, there is no
numerical damping in the Newmark-$\beta$ method.
We observed high-frequency noise in the displacement velocity $\dot{\ww}$ 
calculated with~\Cref{eq:newmark_beta_method}, which is
an issue that can arise when solving problems with prescribed displacement
boundary conditions~\cite{bathe2012insight}. Therefore,
to ensure smoothness of $\dot{\ww}$ before
applying it as a boundary condition in the fluid model,
we computed $\dot{\ww}$ by Fourier transformation
of the displacement $\ww$ and differentiation in the frequency space.

The applied displacement boundary conditions are periodic in time,
with a period equal to one cardiac cycle (one second).
To achieve periodicity in $\dot{\ww}$, we do the following.
The tissue deformation equations \eqref{eq:strong_linear_elasticity_1}
are solved for four cardiac cycles, and the displacement velocities computed in the fourth cycle
are used as boundary conditions in the fluid equations.
Starting the system from rest with initial conditions
$\ww(\xx, 0) = \dot{\ww}(\xx, 0) = \ddot{\ww}(\xx, 0) = \bm{0}$
and smoothly transitioning to the full periodic boundary data
is important to achieve stable results, because
an impulsive start could introduce non-physical oscillations. 
To ensure this smooth start, a cosine window is applied
to the boundary condition magnitudes during the first half of
the initial cardiac cycle. Additionally, we perform a ramp-up
of the magnitudes over the first two cardiac cycles.

\subsubsection{Finite element methods for the fluid equations}
We discretize the governing equations of fluid motion,
the Navier-Stokes (\Cref{eq:navier_stokes_mom_eq,eq:continuity_eq})
and the Stokes (\Cref{eq:continuity_eq,eq:unsteady_stokes_eq,eq:stokes_eq})
equations, with $\bm{H}(\mathrm{div}; \DD)$-conforming finite element methods
using Brezzi--Douglas--Marini (BDM) elements for the velocity
and discontinuous Galerkin (DG)
elements for the pressure~\cite{Brezzi1985TwoProblems,Nedelec1986AIR3}.
The $\mathrm{BDM}_k$--$\mathrm{DG}_{k-1}$ elements, with polynomial order $k$,
yield a spatial discretization that conserves mass exactly pointwise.
For the unsteady Stokes equations, we use a weak formulation
based on the one developed in
Herlyng~\emph{et al.}~\cite{Herlyng2025AdvectionDiffusion}.
This weak form was further developed for the Navier-Stokes equations
\cite{Schroeder2018Towards,di2011mathematical}.

For the $\mathrm{BDM}_k$--$\mathrm{DG}_{k-1}$ elements, we use
polynomial order $k=2$ for the human model and $k=1$ for the 
zebrafish model. The velocity component normal to facets is by definition 
continuous for the BDM elements, while tangential continuity is achieved
by a stabilization technique~\cite{Hong2016AEquations}. 
For the Navier-Stokes equations, the convective fluxes both on facets
internal to the mesh and on open boundaries are stabilized with
an upwinding technique~\cite{Schroeder2018Towards,
di2011mathematical}. The ventricular wall deformations
are incorporated by formulating the governing fluid equations
in an Arbitrary Lagrangian Eulerian framework~\cite{Scovazzi2007LectureNotes,Duarte2004ALE}.
In terms of boundary conditions, the cilia tangential traction
boundary conditions and traction-free conditions are imposed weakly.
The ventricular wall displacement velocity is imposed strongly.
We use a Nitsche method to impose the choroid plexus flux boundary
condition weakly~\cite{Nitsche1971UberSind, benzaken2024constructing}.

In time, the fluid equations are discretized with the first-order, implicit
Euler scheme using a time step $\Delta t = 0.001$ s.
Since we use an implicit time discretization and solve the fluid equations
directly, an initial condition is only needed
for the velocity field, and not the pressure.
When solving the Stokes equations, the fluid is 
initially at rest: $\uu_0(\xx)=\uu(\xx, 0)=\bm{0}$.
For the Navier-Stokes equations, we solve the Stokes equations
(with $\uu_0(\xx)=\bm{0}$) and use the solution
$\uu_{\mathrm{Stokes}}(\xx)$ as initial condition;
$\uu_0(\xx)=\uu_{\mathrm{Stokes}}(\xx)$. 
Moreover, when solving the Navier-Stokes equations the convective term
is linearized as $(\uu^{n+1}\cdot\nabla)\uu^{n+1} \approx (\uu^n\cdot\nabla)\uu^{n+1}$,
where $n$ denotes time step.
To achieve periodic flow in the unsteady simulations, the fluid equations
are solved for two cardiac cycles and the velocities $\uu(\xx, t)$
of the second cycle are used to calculate FTLE fields.
When reporting results, the start of this second cycle is defined as $t=0$.

\subsubsection{Finite-time Lyapunov exponent field computation}
To compute finite-time Lyapunov exponent (FTLE) fields,
we initiate a Cartesian grid of fluid particles at a release
time $t_0$ and compute the flow map $\phi_{t_0}^{t_0+T}$
by solving \Cref{eq:integrate_dx_dt} with a fourth-order Runge-Kutta 
time stepping scheme. The time step size $\Delta t = 0.001$ s
was chosen to satisfy the Courant-Friedrichs-Lewy condition
$\Delta t \leq C h_{\mathrm{min}}/u_{\mathrm{max}}$,
where $C=0.5$, $h_{\mathrm{min}}$ is the minimal edge length
of the mesh, and $u_{\mathrm{max}}$ is
the maximum velocity magnitude. Further details on the
FTLE computation algorithm is provided in previous work~\cite{shadden2010computational}.
Because the FTLE computation algorithm interpolates velocity
based on values in the mesh nodes, the finite element velocity
approximations are projected from $\mathrm{BDM}_k$ into
$\mathrm{CG}_k$ ($k=2$ and $k=1$ for the human and zebrafish model, respectively)
and finally evaluated in a $\mathrm{CG}_1$ space.

\subsubsection{Software implementation}
The FTLE field computations are performed with a 
framework that is implemented in C.
The finite element methods for the fluid and tissue deformation equations
are implemented and solved with DOLFINx~\cite{baratta2023dolfinx}.
A direct solver using MUMPS~\cite{Amestoy2011Mumps} is used to solve the fluid equations.
We solve the tissue equations iteratively with a
Flexible Generalized Minimal Residual (GMRES) method~\cite{fgmres} 
preconditioned with the algebraic multigrid method
BoomerAMG~\cite{yang2002boomeramg} from the hypre library~\cite{hypre}. 
Both the FTLE computation software and the finite element method software
are openly available and archived on Zenodo~\cite{zenodoArchive}.

\section{Results}
\subsection{Pulsatile brain tissue deformation leads to pulsatile CSF flow in the human brain ventricles}
\begin{figure}
    \centering
    \includegraphics[width=\textwidth]{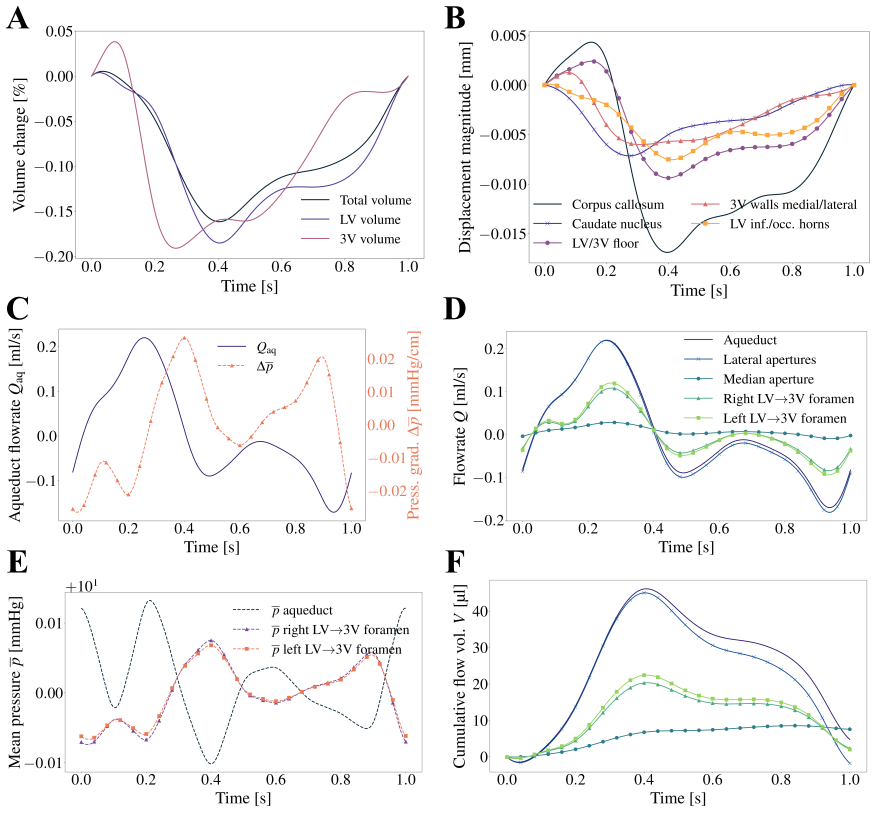}
    \caption{
    \textbf{A.} Volume dynamics of the whole ventricular system, 
    the third ventricle (3V) and the lateral ventricles (LV) over a cardiac cycle.
    \textbf{B.} Illustration of the displacement boundary conditions
    $\bm{g}(\xx, t)$ over a cardiac cycle
    (defined in \Cref{subsubsec:deformation}). The maximum magnitudes of
    $\bm{g}(\xx, t)$ are plotted for the five boundary condition regions:
    corpus callosum; caudate nucleus; lateral and third ventricle floor (LV/3V floor);
    third ventricle medial and lateral walls (3V walls medial/lateral);
    the inferior and occipital horns of the lateral ventricles (LV inf./occ. horns).
    \textbf{C--F.} Quantities of interest computed over one cardiac cycle:
    Flow rate $Q_{\mathrm{aq}}(t)$ and mean pressure gradient $\Delta\overline{p}(t)$
    in the cerebral aqueduct (\textbf{C}); flow rates $Q(t)$ across cross sections
    in different parts of the ventricular system (\textbf{D});
    mean pressures $\overline{p}(t)$ (\textbf{E}); 
    cumulative flow volumes $V(t)$ (\textbf{F}).
    Common legend for panels \textbf{D} and \textbf{F}.
    }
    \label{fig:flow_and_deformation_results}
\end{figure}

When solving the tissue deformation equations with the applied
experimental displacement data, during systole there is an immediate expansion
of the third ventricle (3V) in concert with a cephalic movement of the 3V
and the corpus callosum that lies superior to the lateral ventricles
(\Cref{fig:flow_and_deformation_results}A, B).
This is followed by a contraction of the whole
ventricular system together with a caudal displacement of
the brain ventricles. When solving the
Navier-Stokes equations (\Cref{eq:navier_stokes_mom_eq}) with the
DSC model (deformations, secretion, cilia), the contraction and
caudal displacement leads to a rapid increase in caudal CSF flow
through the cerebral aqueduct and interventricular foramina
(IVF, \Cref{fig:flow_and_deformation_results}C, D).
In the aqueduct, the flow rate peaks at
0.22 ml/s after 26\% of the cardiac cycle (\Cref{fig:flow_and_deformation_results}C).
This is slightly earlier than the greatest
volume displacement of the ventricular system
in terms of magnitude; a volume change of $-0.16\%$ (contraction)
that occurs 40\% throughout the cardiac cycle (\Cref{fig:flow_and_deformation_results}A).
The greatest positive volume displacement (expansion) occurs right after
the onset of the cardiac cycle, constituting 0.01\%.
The lateral ventricles (LV) and the 3V are the main contributors to the volume
change of the ventricular system. In comparison, the fourth ventricle (4V) volume
changes are approximately a thousandth of the total volume change. 
The maximum displacement magnitude of 0.046 mm 
almost coincides with the largest volume displacement ($t=0.41$ s),
and was observed superior to the lateral ventricles on the right side
($\xx = (15.7, -11.2, 29.9)$ mm).

When the ventricles relax during diastole (\Cref{fig:flow_and_deformation_results}A, B),
the CSF flow reverses and flows cephalic to fill the ventricles
and the aqueduct flow rate goes through a local minimum, increases slightly,
then decreases again and bottoms out at $-0.17$ ml/s around
94\% through the cycle (\Cref{fig:flow_and_deformation_results}C).
These flow rate dynamics are similar in the IVF,
and the total flow rate is evenly distributed through both of these foramina
(\Cref{fig:flow_and_deformation_results}D).
Finally, flow reverses when transitioning 
into the next cardiac cycle. Over the cardiac cycle,
the flow rate magnitudes through the aqueduct,
the lateral apertures, and both the IVF
are all very similar (\Cref{fig:flow_and_deformation_results}D).
These regions all have an order of magnitude greater
flow rates than the flow rates through the median aperture.
The maximum CSF velocity magnitude of caudally directed flow is 6.8 cm/s,
which is located in the cerebral aqueduct during systole at 26\% of the cardiac cycle.
For cephalic flow, the minimum velocity is $-4.4$ cm/s in the aqueduct at the end of diastole
after 94\% of the cardiac cycle. The maximum velocity magnitude corresponds to
a Reynolds number of 319, indicating laminar flow.

The gradient of the mean pressure (cross-sectional average) $\Delta\overline{p}$
along the cerebral aqueduct varies inversely with
the flow rate behavior (\Cref{fig:flow_and_deformation_results}C),
with a peak-to-through range of 0.053 mmHg/cm. The mean pressure
at the superior end of the aqueduct oscillates over the cardiac cycle,
going through three local minima and maxima (\Cref{fig:flow_and_deformation_results}E).
The mean pressures in the IVF vary inversely with the aqueduct pressure, and the
mean pressures in the two foramina are similar in magnitude.
Over a cardiac cycle the stroke volume, which is the maximum of the
cumulative flow volume, in the cerebral aqueduct was
46.1 $\mu$l (\Cref{fig:flow_and_deformation_results}F).
Most of the displaced CSF volume crossing the domain boundary
is through the lateral apertures with a stroke volume of 45 $\mu$l,
with less fluid flowing in and out through the median aperture
(stroke volume 8.6 $\mu$l).
The flow volumes through the interventricular foramina connecting the lateral and third 
ventricles are evenly distributed between the two foramina (\Cref{fig:flow_and_deformation_results}D).

\subsection{Flow fields predicted by the Stokes and Navier-Stokes equations 
are similar macroscopically, but differ locally}
Solving the unsteady Stokes equations (\Cref{eq:unsteady_stokes_eq})
yields CSF flow that is qualitatively similar to the
Navier-Stokes flow described in the previous section in terms of the systole and
diastole trends. The Stokes approximation has the same aqueductal stroke volume (46.1 $\mu$l),
but the displaced flow volumes through the lateral apertures (47 $\mu$l) and the
median aperture (4.9 $\mu$l) differ from the Navier-Stokes solution,
especially the median aperture stroke volume, which is 43\%
lower for the Stokes approximation. Although predicted maximum and minimum flow rates and
velocities occur at virtually the same time instants,
the Stokes equations predict lower maximum and minimum velocity magnitudes
(6.0 and $-4.6$ cm/s, respectively). The 12\% lower maximum velocity
magnitude leads to a lower maximum Reynolds number of 282.
The peak-to-nadir pressure gradient equals 0.054 mmHg/cm, slightly higher than that predicted
by the Navier-Stokes equations.

\begin{figure}
    \centering
    \includegraphics[width=\textwidth]{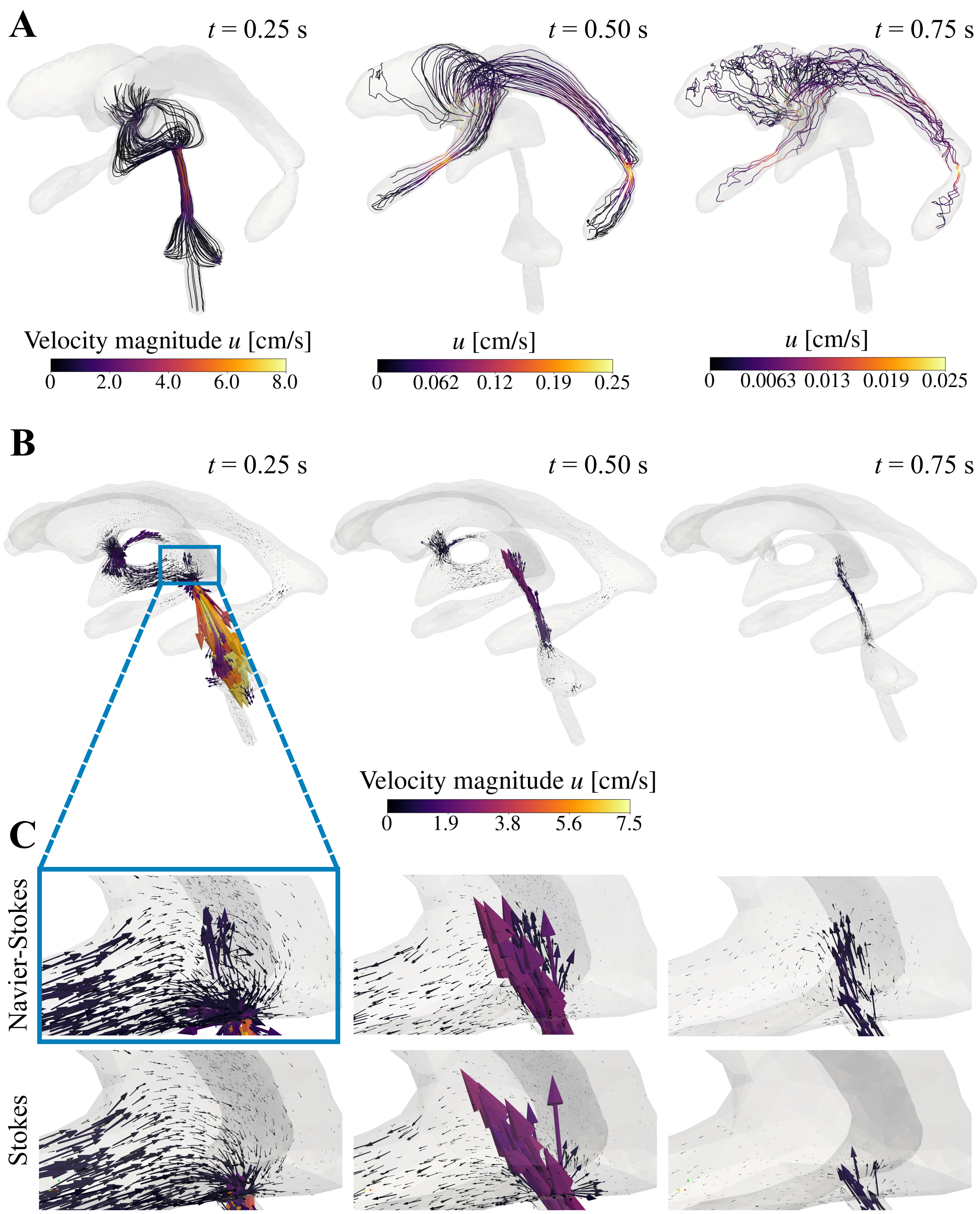}
    \caption{\textbf{A.} Forward-in-time integrated streamlines of particles seeded
    in the interventricular foramina at times $t=0.25, 0.50, 0.75$ s, corresponding
    to systole, mid-cardiac cycle, and diastole, for the Navier-Stokes velocity
    approximation. Note the distinct color bar limit in each plot. 
    \textbf{B.} Navier-Stokes velocity vectors at times $t=0.25, 0.50, 0.75$ s in
    the human brain ventricles with zoomed-in view of  (bottom row).
    \textbf{C.} Zoomed-in view (of the box region shown in the left-most
    subfigure of panel \textbf{B}) of velocity vectors superior to
    the cerebral aqueduct entrance, both for the Navier-Stokes and
    the Stokes solution, to highlight local differences in flow features.
    }
    \label{fig:3d_flow_results}
\end{figure}

Streamlines generated by integration forward in time at a given time
instant with the Navier-Stokes velocity demonstrates that CSF flow is more directional at certain time
instants, while having more sporadic directionality at other times (\Cref{fig:3d_flow_results}A).
Velocity vectors clearly show how the principle of mass conservation results in highest velocities
through narrow sections like the cerebral aqueduct and interventricular foramina
(\Cref{fig:3d_flow_results}B).
Velocities in larger compartments, like the lateral ventricles, are several orders 
of magnitude smaller. For example, during systole the velocities in the aqueduct
reach 7.5 cm/s, while in the lateral ventricles the velocities are in the order of
0.1 cm/s. A close-up inside the third ventricle reveals disparate flow features for 
the Navier-Stokes and Stokes approximations (\Cref{fig:3d_flow_results}C).
The Navier-Stokes CSF velocity vectors have vortical patterns
indicating a jet forming right above the entrance to
the cerebral aqueduct (\Cref{fig:3d_flow_results}C).
These vortical patterns are absent in the Stokes solution.

\subsection{Disregarding choroid plexus secretion or motile cilia lead to small changes in flow quantities}
When comparing results for the three flow model versions (DSC, DC, DS), most of the computed flow quantities are similar.
The aqueductal and lateral aperture stroke volumes are the same when excluding cilia (DS) as for the full model (DSC).
Removing secretion while keeping cilia (DC model), there is slightly less flow with a 4\% decrease in the aqueductal
and lateral aperture stroke volumes when compared to DSC, with a similar decrease in the interventricular foramina stroke volumes.
The median aperture stroke volume is less for both DS and DC compared to DSC, with 11\% and 12\% decreases for DS and DC,
respectively. The pressure gradient peak-to-nadir is unaltered for both DC and DS.
The maximal and minimal velocity magnitudes and flow rates change
by less than 4\% and 3\% for DC and DS, respectively.

\subsection{Motile cilia generate rotational flow in embryonic zebrafish brain ventricles}
\begin{figure}
    \centering
    \includegraphics[width=\textwidth]{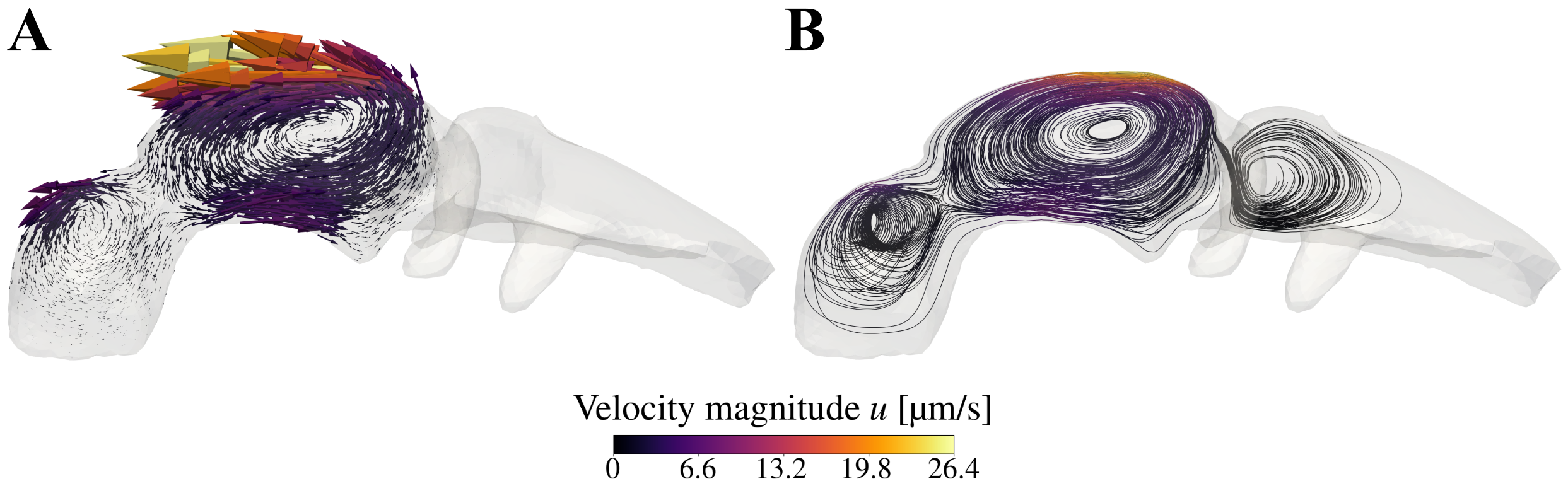}
    \caption{\textbf{A.} Velocity vectors scaled by magnitude.
    \textbf{B.} Streamlines of particles integrated forward in time,
    with particles seeded in the anterior,
    middle and posterior ventricles.
    Common color bar for \textbf{A} and \textbf{B}.
    }
    \label{fig:zfish_flow_results}
\end{figure}

Solving the (steady-state) Stokes equations in the zebrafish brain ventricles, 
the motile cilia generate a rotational CSF flow with vortical
structures confined to the different ventricular cavities (\Cref{fig:zfish_flow_results}, \textbf{a}).
As a result of the applied tangential
traction being constant and the geometry being closed, this flow field is a steady-state solution.
The maximum velocity magnitude is 26.4 $\mu$m/s.
Streamlines of particles integrated forward in time show rotational particle trajectories
(\Cref{fig:zfish_flow_results}, \textbf{B}). In \Cref{fig:zfish_flow_results}, particles
are seeded in the anterior, middle and posterior ventricles, and the three rotational
structures originate from each of the particle seeding locations. Only seeding particles
in e.g. the anterior ventricle would result in streamlines almost exclusively confined 
to the anterior ventricle, with only few particles crossing through the interventricular ducts.

\subsection{Finite-time Lyapunov exponent fields demonstrate prominent
flow structures in the human brain ventricles}
Distinct flow features (such as jets) have in previous computational studies
been observed in the third ventricle at the cerebral aqueduct
entrance~\cite{Kurtcuoglu2007ComputationalSylvius}.
We therefore investigate structures in the Finite-time Lyapunov exponent
(FTLE) fields near this entrance and in the interventricular foramina.
The cerebral aqueduct and the interventricular foramina
are among the narrowest sections of the ventricular system, where flow
gradients increase as a result of the principle of mass conservation.

\begin{figure}
    \centering
    \includegraphics[width=\textwidth]{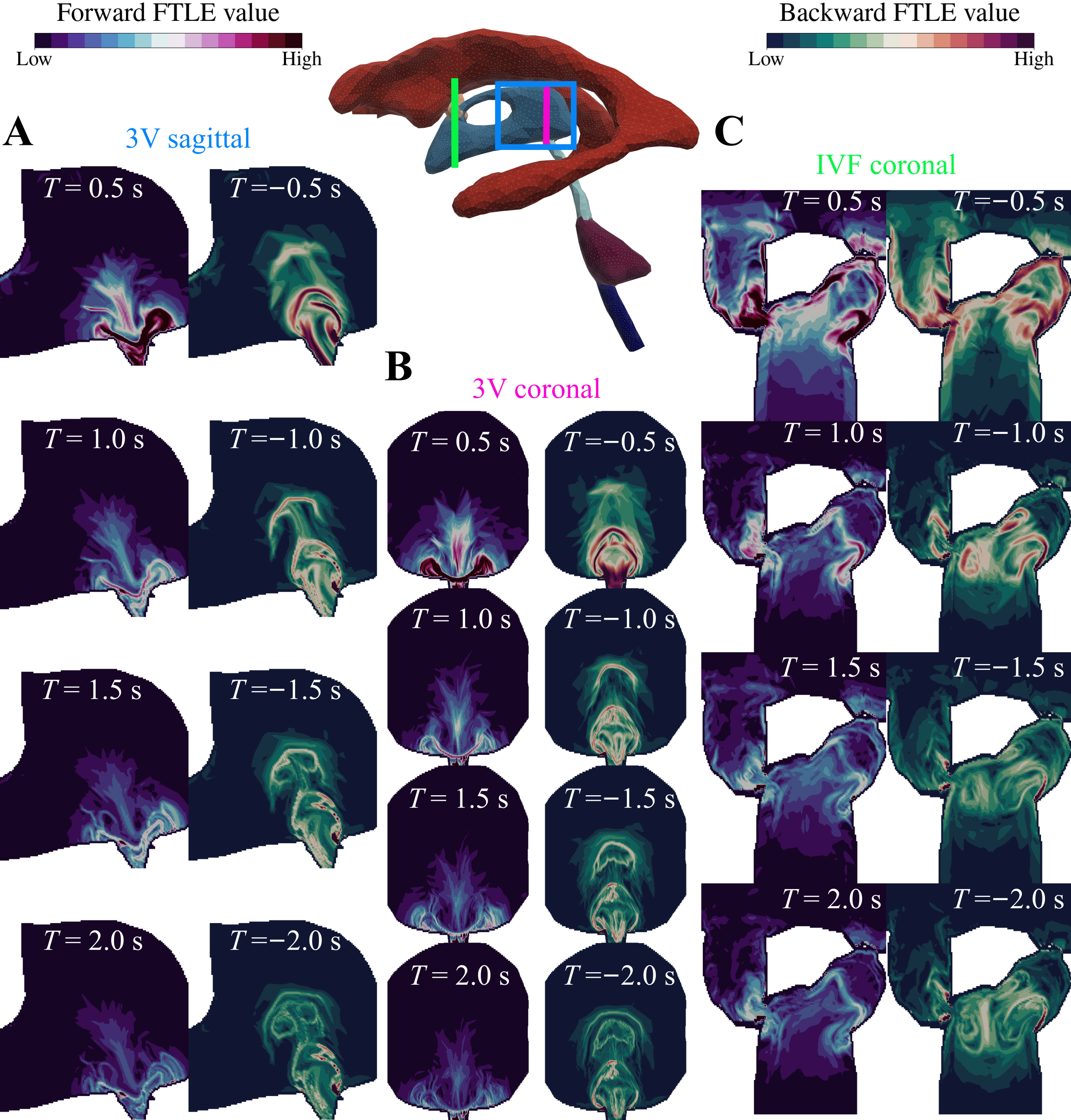}
    \caption{Slices of FTLE fields computed with initial release time
    $t_0=0$ and integration times $T = \pm\{0.5, 1.0, 1.5, 2.0\}$ s
    in the human brain ventricles. The FTLE fields
    were computed in two box regions: (i) the posterior end of the third
    ventricle, including the superior most part of the aqueduct, and
    (ii) the anterior part of the lateral ventricles, including
    the interventricular foramina. Sagittal slices in the 3V (\textbf{A})
    and coronal slices in the 3V (\textbf{B}) and the IVF (\textbf{C})
    of these FTLE fields are displayed. Within each slice-specific column,
    the left and right columns display forward- and backward-time FTLE
    fields, respectively.    
    The lower color bar limit is zero in all slices,
    while the higher color bar limit is
    an FTLE value of 2.5 and 5.0 for the IVF and 3V slices, respectively.}
    \label{fig:FTLE_human1}
\end{figure}

FTLE fields computed using the solution to the Navier-Stokes
equations (\Cref{eq:navier_stokes_mom_eq}) with varying integration time $T$
show the emergence of a jet and the generation of a vortex ring
at the cephalic entrance to the aqueduct (\Cref{fig:FTLE_human1}A, B).
The leading edge of the vortex ring is identified by ridges in the backward-in-time
FTLE fields, ridges that develop with increasing $T$.
In the forward-in-time FTLE field, there is a well-defined ridge
for $T=1.0$ s (one cardiac cycle) across the entrance to
and the cerebral aqueduct (\Cref{fig:FTLE_human1}A, B).

There are also clearly visible ridges that define lobe-like structures
in the FTLE fields in the interventricular foramina (IVF) (\Cref{fig:FTLE_human1}C).
In the backward FTLE fields there develops a vortex ring that is ejected
caudally. In the bulk of the lateral and third ventricles,
there are no clear structures in the FTLE fields.
We observe clearer structures as the integration time increases through
a couple of cardiac cycles. For higher integration times, the structures become
smeared out with diminishing FTLE values. 

\begin{figure}
    \centering
    \includegraphics[width=\textwidth]{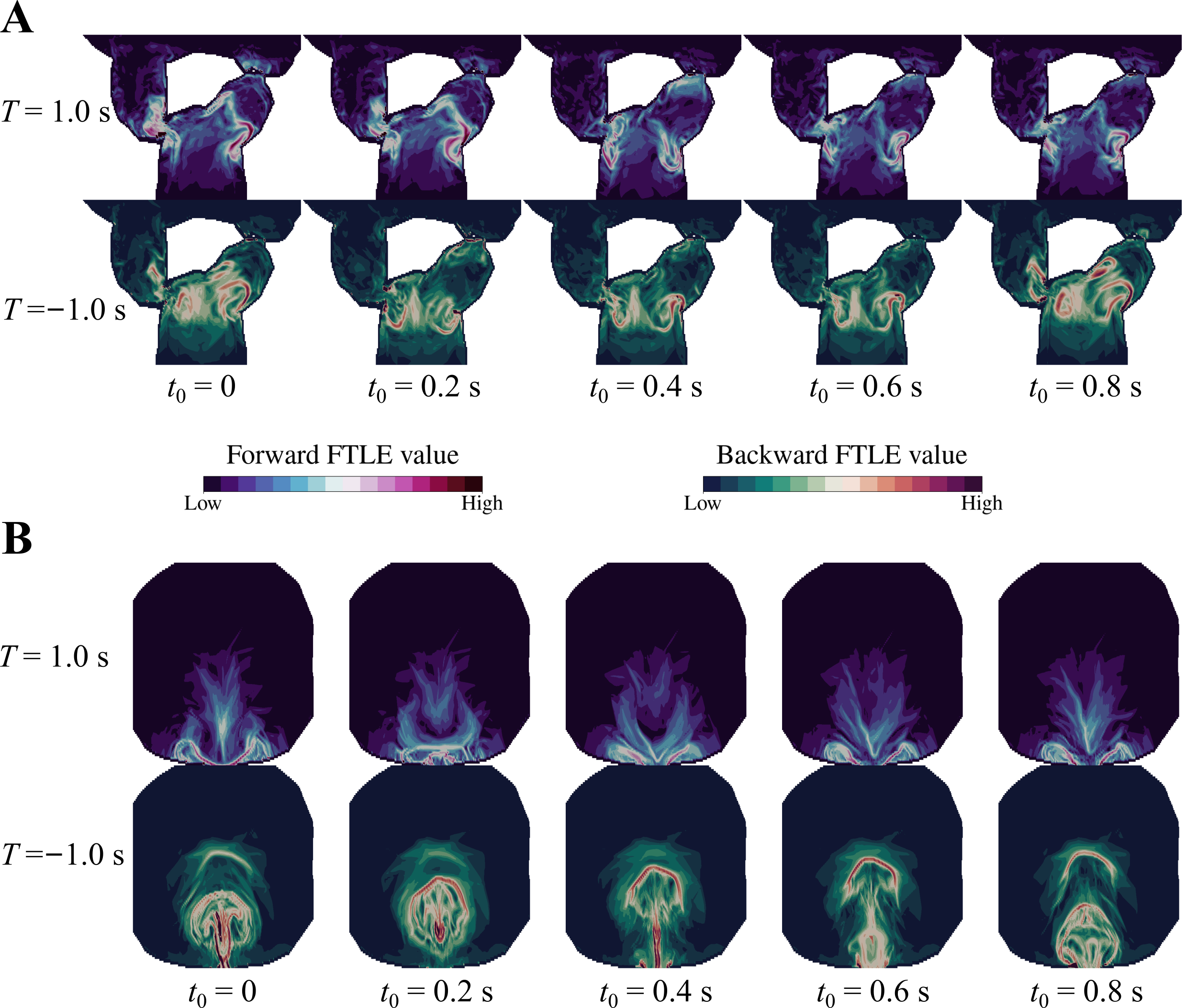}
    \caption{Coronal slices of FTLE fields in the human brain ventricles 
    computed with integration times $T = \pm 1.0$ s
    (one cardiac cycle). The FTLE fields are computed with varying
    initial release times $t_0$.
    The top two rows are slices taken in
    the IVF, and of these two rows the top and bottom display
    forward- and backward-time FTLE fields, respectively.
    The bottom two rows are sagittal slices taken in the 3V (forward- and 
    backward-time FTLE displayed at the top and bottom).
    }
    \label{fig:FTLE_human2}
\end{figure}

For an integration time of one cardiac cycle ($T=1.0$ s), FTLE fields computed
with varying initial release times $t_0$ show how the FTLE fields vary
in time (\Cref{fig:FTLE_human2}). In the IVF,
the attracting structures reverse their orientation over the course of
one cardiac cycle (one second, \Cref{fig:FTLE_human2}A).
In the 3V, the vortex formed by the
jet near the aqueduct entrance detaches from the entrance 
during early systole (\Cref{fig:FTLE_human2}B).
A new vortex is forming during the transition to diastole,
and a new vortex clearly forms during the transition to the next cardiac cycle.

\subsection{When neglecting fluid inertia, the model fails to resolve important Lagrangian structures}
\begin{figure}
    \centering
    \includegraphics[width=0.75\textwidth]{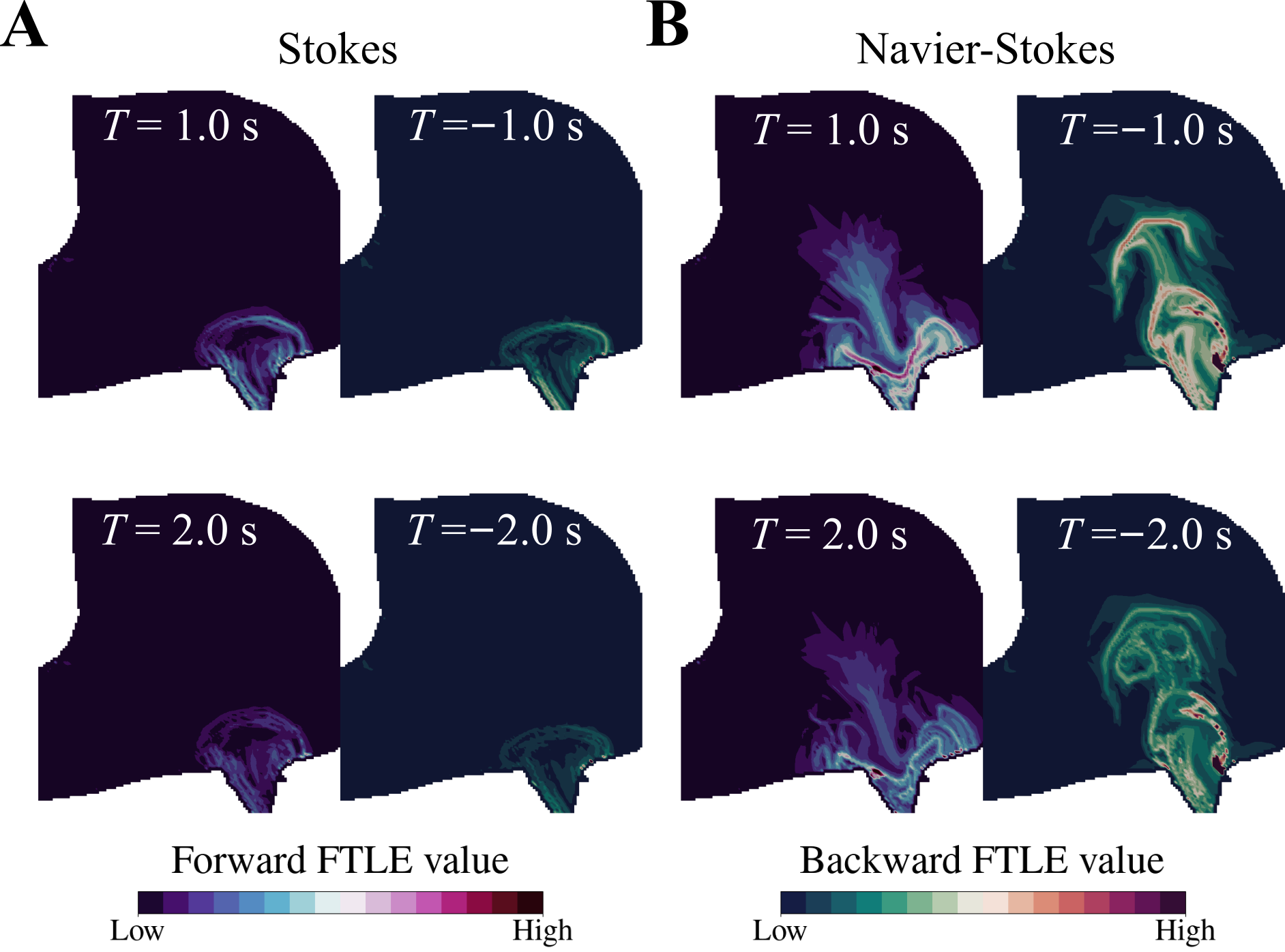}
    \caption{Sagittal slices of FTLE fields computed in the human brain ventricles
    using the unsteady Stokes equations (\textbf{A})
    and the Navier-Stokes equations (\textbf{B})
    with initial particle release time $t_0=0$ and integration times 
    $T = \pm\{1.0, 2.0\}$ s. 
    In both subpanels, forward-time FTLE are displayed in the left
    column while backward-time FTLE are displayed in the right column.
    }
    \label{fig:FTLE_human3}
\end{figure}

When computing FTLE fields with the unsteady Stokes equations (\Cref{eq:unsteady_stokes_eq}),
the jet structure in the third ventricle vanishes (\Cref{fig:FTLE_human3}A).
This differs greatly from the structure
observed in the Navier-Stokes solution (\Cref{fig:FTLE_human3}B).
In the backward FTLE fields, the lobes that form and over time
propagate away from the aqueduct entrance in the Navier-Stokes solution are not 
present in the Stokes solution's FTLE fields. This indicates a stronger stirring effect
in the Navier-Stokes solution. Similarly, the repelling structures are more
extensive in the Navier-Stokes FTLE fields.

\subsection{Brain tissue deformations are the main contributor to establishing Lagrangian structures}
\begin{figure}
    \centering
    \includegraphics[width=0.75\textwidth]{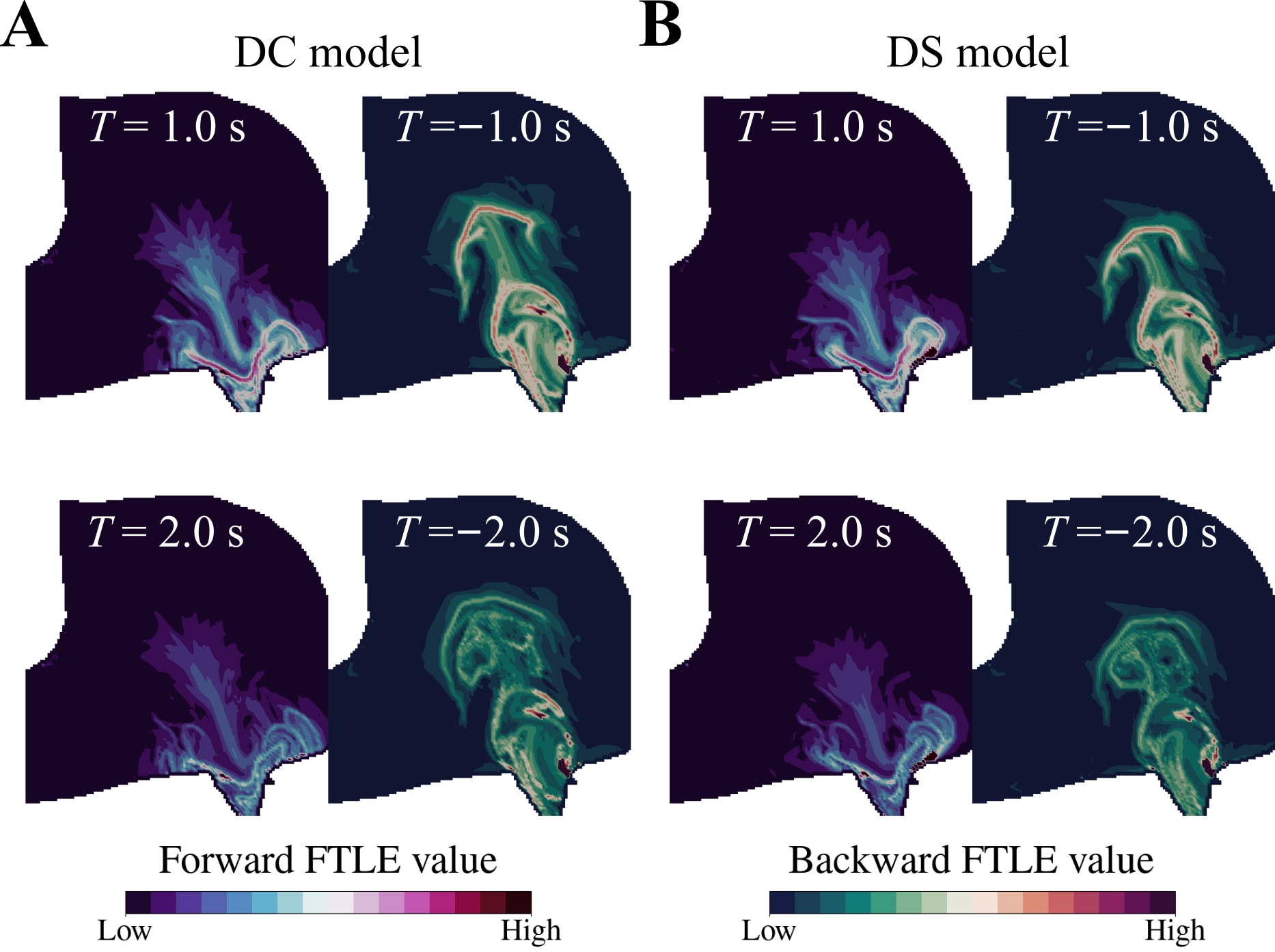}
    \caption{Sagittal slices of FTLE fields computed in the human brain ventricles
    using the DC model (\textbf{A}, no secretion)
    and DS model (\textbf{B}, no cilia)
    with initial particle release time $t_0=0$ and integration times 
    $T = \pm\{1.0, 2.0\}$ s. 
    In both subpanels, forward-time FTLE are displayed in the left
    column while backward-time FTLE are displayed in the right column.
    }
    \label{fig:FTLE_human4}
\end{figure}

The FTLE fields computed in the third ventricle with the DC and DS models
are generally comparable to the DSC results, with the formation of
a vortex ring (\Cref{fig:FTLE_human4}).
When excluding secretion (DC model), the leading edge of the vortex
advances slightly faster (\Cref{fig:FTLE_human4}A).
Meanwhile, excluding cilia (DS model) makes the forward-time FTLE field
less attached to the ventricular walls close to the aqueductal entrance
(\Cref{fig:FTLE_human4}B).

\subsection{FTLE fields demonstrate brain ventricular flow compartmentalization in zebrafish embryos}
\begin{figure}
    \centering
    \includegraphics[width=\textwidth]{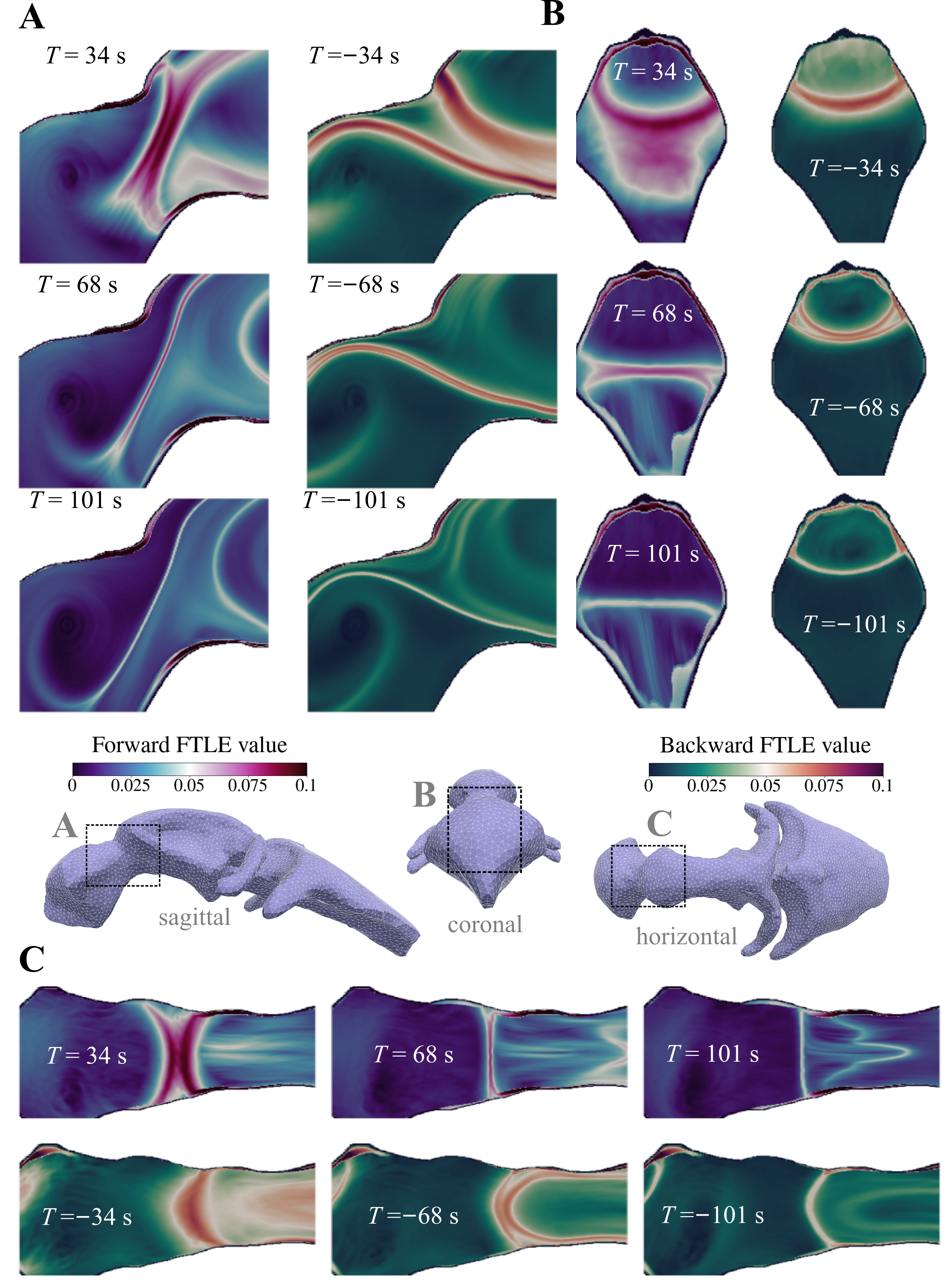}
    \caption{FTLE fields for integration times $T = \pm\{34, 68, 101\}$ s
    in embryonic zebrafish brain ventricles. The FTLE fields
    were computed in a box section of the interventricular duct
    connecting the anterior and middle ventricles.
    Sagittal (\textbf{A}), coronal (\textbf{B})
    and horizontal slices (\textbf{C}) in the center of the box
    are displayed. Common legends for all forward-in-time
    and all backward-in-time fields.}
    \label{fig:FTLE_zfish}
\end{figure}

Experimental and computational studies have demonstrated compartmentalization of
brain ventricular flow in embryonic zebrafish ventricles~\cite{Olstad2019CiliaryDevelopment,
Herlyng2025AdvectionDiffusion}. We therefore investigate structures in the FTLE fields
in the interventricular duct connecting the anterior and middle ventricles to see
if we can observe such flow compartmentalization in the structures defined by the FTLE
fields. As observed in \Cref{fig:FTLE_zfish}, 
there are clear ridges in the FTLE fields defining attracting and repelling LCS.
These repelling and attracting LCS combine to partition regions of distinct
flow dynamics. The LCS confine CSF to respective ventricles,
demonstrating compartmentalization of flow.
The integration time $T$ required to reveal these structures is more than one
order of magnitude larger than the $T$ values used for the human ventricles,
a result of the lower velocities in the zebrafish brain ventricles.

\section{Discussion}

\subsection{Lagrangian structures are present in the ventricular system}
Both in the zebrafish and the human model, we find ridges in the finite-time
Lyapunov exponent fields (FTLE) that indicate the presence
of Lagrangian coherent structures (LCS), demonstrating organization
of the CSF flow. In the human ventricular system, the jet observed in 
the third ventricle (\Cref{fig:FTLE_human1}) is consistent with previous computational
studies~\cite{Kurtcuoglu2007ComputationalSylvius}. The unsteady FTLE
fields (\Cref{fig:FTLE_human2}) indicate that the jet creates a vortex ring every cardiac cycle.
As characterized in previous studies, this vortex ring indicates
strong stretching and folding of the CSF,
contributing to mixing and stirring~\cite{shadden2006lagrangian, shadden2007transport}. 
The vortex ring also traps fluid, thus the presence
of the ring suggests some CSF may occupy the third ventricle for
relatively longer times.

For $T=1.0$ s (one cardiac cycle) we identified a repelling structure 
in the third ventricle. This repelling structure separates the third
ventricle from the aqueduct, which means
that once fluid has reached the entrance to the aqueduct, it is quickly
expelled through the aqueduct. Meanwhile, fluid inside the third ventricle
can stay there for a longer time without exiting through the aqueduct.
In the interventricular foramina, there are also
clear ridges in the FTLE field with $T=1.0$ s (\Cref{fig:FTLE_human1}),
indicating regions of high stretching that partition flow
with different advective transport times. The lobes that develop
have a structure that could trap fluid close to the ventricular walls.

In the zebrafish ventricles FTLE fields (\Cref{fig:FTLE_zfish}),
there are clearly defined LCS, both attracting and repelling,
in the interventricular duct separating the anterior and middle ventricles.
Combining the backward- and forward-in-time FTLE fields,
the prominent attracting and repelling structures form a hyperbolic separation point. 
This confines flow to separate parts of the ventricular system.
This finding aligns well with previous studies that have demonstrated flow
compartmentalization in embryonic zebrafish
brain ventricles~\cite{Olstad2019CiliaryDevelopment,
Herlyng2025AdvectionDiffusion}.

\subsection{Accounting for fluid inertia is important to resolve Lagrangian structures,
but less important for macroscopic flow analysis}
The results we present indicate that when simulating CSF flow
in human ventricles at Reynolds numbers in the lower hundreds,
accounting for inertial effects through
solving the full Navier-Stokes equations is important when investigating
advective transport by the CSF flow and analyzing local flow features.
We draw this conclusion based on the significant differences in the structures
present in the FTLE fields (\Cref{fig:FTLE_human3}) and the
differences observed in local flow features (\Cref{fig:3d_flow_results})
between the Navier-Stokes and Stokes solutions. The lobe structures and
vortex rings that form in the Navier-Stokes FTLE fields
indicate stronger stirring and mixing of
CSF than in the CSF flow approximated with the unsteady Stokes equations.

In terms of computing macroscopic or integrated quantities like stroke volume,
however, the unsteady Stokes equations seem sufficiently accurate, showing little discrepancy with
the Navier-Stokes solution. As such, the Stokes equations can be
utilized as a low-cost alternative for fast predictions
in computational frameworks where integrated quantities are needed.
We note that several previous computational studies report maximal aqueductal
CSF velocities on a mm/s scale~\cite{enzmann1993cerebrospinal, enzmann1991normal,
Causemann2022HumanFramework, Sweetman2011ThreeDimensional, howden2008three, wagshul2011pulsating,
greitz1993pulsatile, sweetman2011cerebrospinal}, which is ten times lower than the
cm/s scale reported here and in several other works
\cite{kim1999quantitative, nitz1992flow, Kurtcuoglu2007ComputationalSylvius}. 
This ten-fold reduction in maximal velocity would reduce the Reynolds number
by a factor ten, yielding for our model Reynolds numbers in the tens instead of the hundreds.
Although inertial forces would still dominate viscous forces at 
these numbers, investigating the effects on the discrepancy in the advective transport
between the Stokes and Navier-Stokes solutions would be interesting to address in future work.

\subsection{Tissue deformations alter Lagrangian structures more significantly
than motile cilia and CSF secretion}
Disregarding motile cilia (DS model) did not alter the flow in the human brain ventricles significantly
in terms of macroscopic flow quantities like stroke volume. 
There were however minor differences in the FTLE field structures computed with the DS model compared
to both DC and DSC, with the lobe structures near the aqueduct
entrance more attached to the 3V wall when including cilia (DC/DSC),
indicating that cilia alter the near-wall flow (\Cref{fig:FTLE_human4}).
This finding compares well with the computational study
by Siyahhan~\emph{et al.}~\cite{Siyahhan2014FlowVentricles},
whose results showed only a minor effect of cilia on near-wall flow in the lateral
ventricles. On the other hand, in the model of Yoshida~\emph{et al.}~\cite{Yoshida2022EffectVentricles} 
cilia had a significant effect on mixing. We observed that the cilia model parameters
strongly influenced the effect of cilia on the flow (see \Cref{subsec:limitations}),
and we are of the opinion that any further insights on the effect of cilia in
brain ventricular flow and transport would heavily benefit from more experimental data
on cilia parameters in the human brain ventricles.
This contrasts the case for the embryonic zebrafish brain, where the model parameters
are more comprehensively resolved, and previous work strongly suggests cilia are
the main drivers of flow~\cite{Olstad2019CiliaryDevelopment, Herlyng2025AdvectionDiffusion}.
The greater effect of cilia in zebrafish ventricles is a direct result of the smaller
length scale of the ventricular system of embryonic zebrafish
as compared to adult humans. Cilia may play a similarly important
role in smaller confined spaces of the human ventricular system,
such as in the inferior and occipital horns of the lateral ventricles.

The constant-in-time CSF secretion we imposed had little impact on the
computed flow quantities, apart from generating a net outward flow from the
ventricular system. In terms of local flow features,
these were largely unaffected when disregarding CSF secretion (DC model)
compared to the DSC model. In the FTLE fields, we observe a small
difference in the location of the leading edge of the vortex ring
when disregarding secretion (\Cref{fig:FTLE_human4}).
This may result from the secretion increasing the total CSF volume
flowing through the system, leading to a faster development of the vortex ring.

\subsection{Model validation with experimental data of brain deformations and ventricular CSF flow}
We calibrated the displacement boundary data $\bm{g}(\xx, t)$
to ensure a physiological magnitude of CSF flow
\cite{eide2021direction,wagshul2011pulsating}.
The calibration resulted in a reduction in the magnitude of the displacements,
while retaining their qualitative behavior. With our calibrated displacement data,
the maximum volume change in our model was around 0.15\%.
This is less than that reported by previous research,
where 0.9\%~\cite{zhu2006dynamics} and even as high as 20\%~\cite{lee1989variation}
has been reported. These higher values would produce unrealistically large flow
volumes in our model, and the discrepancies between our model and
experimental reports indicate the absence of another
volume-regulating mechanism in our model. One potential explanation could be
that our model lacks the parts of the CSF flow system downstream
of the ventricles, and models the median aperture and lateral apertures as 
open (traction-free) boundaries. In reality, there is some flow resistance
since CSF downstream of the ventricular system must be displaced
if the ventricular CSF is to be displaced. 

The cephalocaudal displacements we imposed are
well-documented~\cite{Kurtcuoglu2007ComputationalSylvius,
Enzmann1992BrainImaging, greitz1992pulsatile, soellinger2007assessment, soellinger20093d},
but there is uncertainty regarding displacements 
in the medial-lateral direction. Our understanding is
that arterial swelling exerts pressure on the parenchyma and results in
squeezing of the ventricular system, a squeezing which seems to be symmetric
in the horizontal plane, as symmetry characterizes medial-lateral motion
in some experimental reports~\cite{soellinger20093d, abderezaei2021development}.
It has been reported that the periventricular deformations could be related to
deformations in the third and lateral ventricles~\cite{abderezaei2021development}.
We imposed medial-lateral displacements on the third and lateral 
ventricles based on these reported findings, and the fact that our model
was unable to produce physiological CSF flow without these displacements;
when imposing only cephalocaudal displacements, the CSF flow rates were less than
10\% of experimentally reported values. This resonates well with the conclusion of
Kurtcuoglu~\emph{et al.}~\cite{Kurtcuoglu2007ComputationalSylvius},
namely that medial-lateral displacement of the third ventricle is
required to fully explain the observed flow rates of ventricular CSF flow.

We considered the beating direction of cilia to align with the direction of 
CSF production velocities, similar to Siyahhan~\emph{et al.}, who considered
directionality of the cilia forcing aligned
with the wall shear stress of the macroscopic flow simulated in a
production-only setup~\cite{Siyahhan2014FlowVentricles}.
In our setup, with the only free boundaries of the geometry located
in the caudal end of the geometry, the production velocity and thus
the ciliary beating is generally caudally directed. Assessing the impact
of this directionality, and whether e.g. a more cephalic directed beating
influences the results, could be investigated in future work.

\subsection{Limitations}
\label{subsec:limitations}

This modeling work comprises several mechanisms, parameters and
assumptions that are uncertain.
For the tissue deformation equations, we used a damped linear elastic model
where there is an inherent
uncertainty in the material parameters
of brain tissue~\cite{smillie2005hydroelastic}.
Biological soft tissues rarely exhibit Hookean material behavior,
and nonlinear models are presumably most appropriate to
model deformations over longer time spans
\cite{Cheng2010ComputationalModel, Cheng2007UnconfinedCompression,
smillie2005hydroelastic, tully2009coupling, kaczmarek1997hydromechanics}.
We justify our choice of deformation model by the fact that
we seek simply a smooth extension to the rest of the domain boundary
for the displacement measurement boundary condition $\bm{g}(\xx, t)$,
while the ventricles are in reality fluid-filled cavities.
The choice of linear elasticity with linear damping is considered
adequate since we consider deformations on the timescale of a cardiac cycle,
where nonlinear damping effects are negligible.

In terms of the flow model, there are mechanisms of CSF flow that we disregarded.
For example, respiration has an influence~\cite{Vinje2019RespiratoryMeasurements},
while we only modeled pulsatile deformations driven by a cardiac frequency.
We disregarded the respiration partly because cardiovascular pulsations
have been found to be the dominant pulsatile mechanisms
in driving CSF flow~\cite{Soderstrom2025}. Other mechanisms like 
absorption of CSF and neural activity were not modeled because 
of our focus on the impact of cilia and CSF secretion on the
ventricular flow and transport.
Additionally, we comment that our model lacks other
parts of the CSF flow system such
as the spinal canal and the subarachnoid spaces,
a choice made based on our focus on ventricular flow. 

We model CSF secretion as constant in time and evenly distributed
across the choroid plexuses. These assumptions may not perfectly reflect
reality, as there are still debates whether CSF may be produced
elsewhere in the brain~\cite{miyajima2015evaluation}.
For example, Brodal~\cite{Brodal2016TheSystem}
report that in reality 10--30\% of the CSF production
may come from interstitial fluid transported through the ependymal cell
layer of the brain ventricles walls into the ventricles.
A variable-in-time production of CSF and a different distribution of the production
magnitudes between the different choroid plexuses could impact our results. 

As already mentioned, there is also uncertainty around our choice
of Navier slip boundary condition and the associated parameter values.
First, the force per surface area parameter $\tau$ is based on the value used by
Siyahhan~\emph{et al.}~\cite{Siyahhan2014FlowVentricles}, which they obtained by
calibrating CSF flow simulations with rodent CSF velocity data. Because the
ependymal cells lining the human brain ventricular walls
in general are ciliated to a higher degree than in the rodent brain~\cite{spassky2017development},
the ciliary beating might be stronger in human brain ventricles
than the beating in the rodent brain. Second,
the friction coefficient $\alpha=47 \ \mathrm{Pa\cdot s/m}$
is estimated by assuming a slip length equal to the depth of the cilia layer lining
the wall. This was chosen partly by comparing the cilia-generated velocities for
different values of $\alpha$. Setting $\alpha=0$ like in the zebrafish model,
the wall velocities exceeded 1 cm/s, which we considered large compared to measured
values and our current understanding of ventricular flow.
On the other hand, a far higher value $\alpha=1000 \ \mathrm{Pa\cdot s/m}$
was used by Causemann~\emph{et al.}~\cite{Causemann2022HumanFramework}
who modeled the ventricular walls as porous with transport into the neighboring
parenchyma. However, $\alpha=1000 \ \mathrm{Pa\cdot s/m}$
is approaching a no-slip regime. The resulting
particle paths generated would thus likely not compare well with
in vivo data~\cite{stadlbauer2010insight}. We therefore
chose the value of $\alpha=47 \ \mathrm{Pa\cdot s/m}$ as a 
compromise between these two cases. Further testing of values for
both $\tau$ and $\alpha$ may provide further insights on this.

\section{Conclusions and outlook}
We have presented a computational framework for using finite-time Lyapunov exponent (FTLE)
fields to characterize Lagrangian coherent structures in brain ventricles.
The FTLE fields are computed from velocity fields predicted with finite element simulations
of cerebrospinal fluid (CSF) flow. Of the three flow mechanisms we considered, cardiovascular pulsation-related
deformations, choroid plexus secretion, and motile cilia, the deformations are the main contributor to
establishing Lagrangian structures in the brain ventricles. Additionally,
we demonstrated that accounting for fluid inertia
in flow field predictions has little importance when computing integrated flow quantities,
but fluid inertial effects are crucial for capturing Lagrangian coherent structures and thus crucial when
modeling transport and mixing in the (human) brain ventricles. Through comparison with
Eulerian-based flow analysis of CSF flow in embryonic zebrafish brain ventricles, our results further
demonstrate how FTLE fields can be used to characterize prominent flow structures in brain ventricles
and partitioning of the flow.
An interesting path for future work would be to assess how these structures change
with either pathological flow regimes or changes in the ventricular geometry.
Although there are still several unresolved model parameters and assumptions,
this work hopefully inspires future work on Lagrangian analysis of brain ventricular flow and transport.

\subsection*{Acknowledgments}
H. Herlyng thanks Miroslav Kuchta, Marie E. Rognes
and Marius Causemann for helpful model discussions,
and Nathalie Jurisch-Yaksi for
insightful discussions about the ventricular system.
H. Herlyng acknowledges computational support from the
Experimental Infrastructure for Exploration of Exascale
Computing (eX3). H. Herlyng acknowledges support from
the Research Council of Norway through 
the Simula Berkeley Education and Research Collaboration
(SIMBER, INTPART project number 322312).


\bibliography{references}{}
\bibliographystyle{unsrturl}

\newpage
\appendix

\section{Model verification}
Correctness of the zebrafish CSF flow model implementation
was verified in previous work~\cite{Herlyng2025AdvectionDiffusion},
thus we only present verification analysis of the human brain model here.
Mesh and polynomial order refinement studies for the human
model were performed both for the fluid and the tissue deformation equations.

\subsection{Finite element methods for the fluid equations}
Correctness of the discretization schemes on the human brain ventricles geometry
were verified using the Method of Manufactured Solutions~\cite{Roache2001CodeSolutions}.
Convergence properties of the fluid discretization scheme were assessed with two
model problems. To assess spatial convergence, we solved a channel flow
problem, observing linear convergence in both the velocity error measured
in the $\bm{H}^1$ semi-norm and the pressure error measured in the $L^2$ norm
when employing $\mathrm{BDM}_1$--$\mathrm{DG}_0$ elements.
Temporal convergence was assessed with a Taylor vortex
problem~\cite{Karniadakis2005Verification}, where the first-order
accuracy of the backward Euler scheme was confirmed.

On the human brain ventricles mesh, convergence was assessed for computed quantities
of interest like stroke volumes and maximum and minimum velocity
magnitudes, by computing these with three different finite element
polynomial orders. Comparison of these computations was also carried out on the coarse and
the standard human mesh, where consistency in results was observed.
We also assessed cardiac cycle independency, and observed a negligible error
between simulating for 2 or 3 cardiac cycles, justifying our choice of 2 cycles
in the main body of the text. 

\subsection{Finite element method for the elasticity equations}
We refine element polynomial degree on a single mesh.
We also refine mesh with a single element degree.
We consider three versions of the ventricles mesh,
with a total of 51 016 (maximal 0.73 cm, minimal 0.049 cm), 408 128 (maximal 0.40 cm, minimal 0.023 cm),
and 3 265 024 (maximal 0.22 cm, minimal 0.010 cm) tetrahedral cells.
We look at convergence in the maximum displacement magnitude, the
displacement in a point on the traction-free boundary, and
the kinetic and elastic energies.

Cardiac cycle independency was assessed. We simulated flow for
2 cardiac cycles and deformation for 4 cardiac cycles. Simulating flow for 4
instead of 2 cardiac cycles showed made virtually no change to most of the computed flow
quantities, with the biggest discrepancy being a 1.5\% change in the max-to-min pressure
gradient, verifying periodicity. The displacement results were also unchanged when simulating
for more than 4 cardiac cycles. These results justify our choice.

We verified the implementation of the linear elasticity equations
by numerically solving the equations with quadratic basis functions. The errors of the displacement
approximation measured in the $\bm{L}^2$ norm and
$\bm{H}^1$ semi-norm converged at the optimal rates of three
and two, respectively. In time, second-order accuracy of the 
Newmark-$\beta$ method was confirmed.

\begin{figure}
    \centering
    \includegraphics[width=0.8\textwidth]{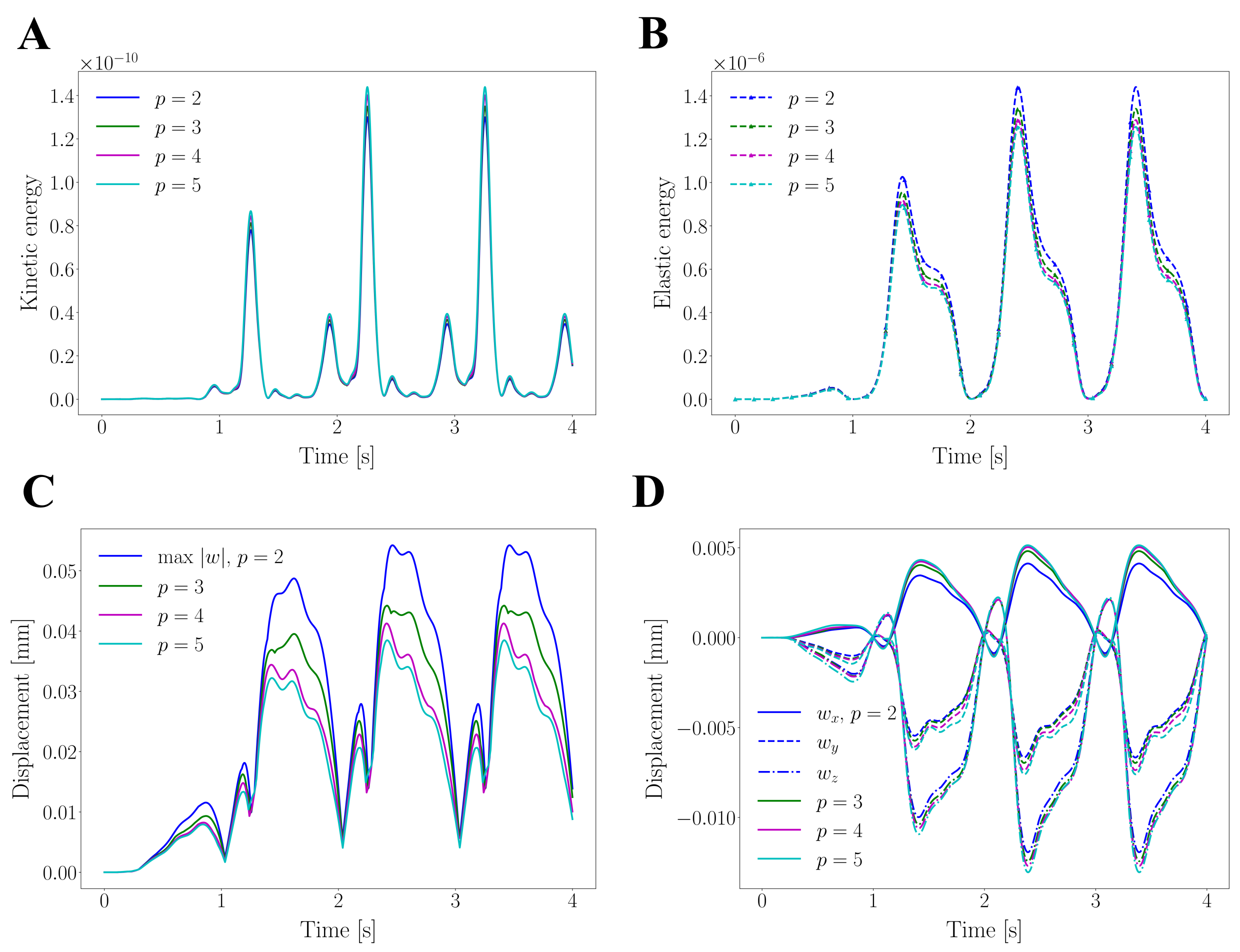}
    \caption{Convergence under element degree ($p$) refinement for the linear elastodynamics model
    on the standard mesh (mesh 1).}
    \label{fig:verification_deformation_p}
\end{figure}
\begin{figure}
    \centering
    \includegraphics[width=0.8\textwidth]{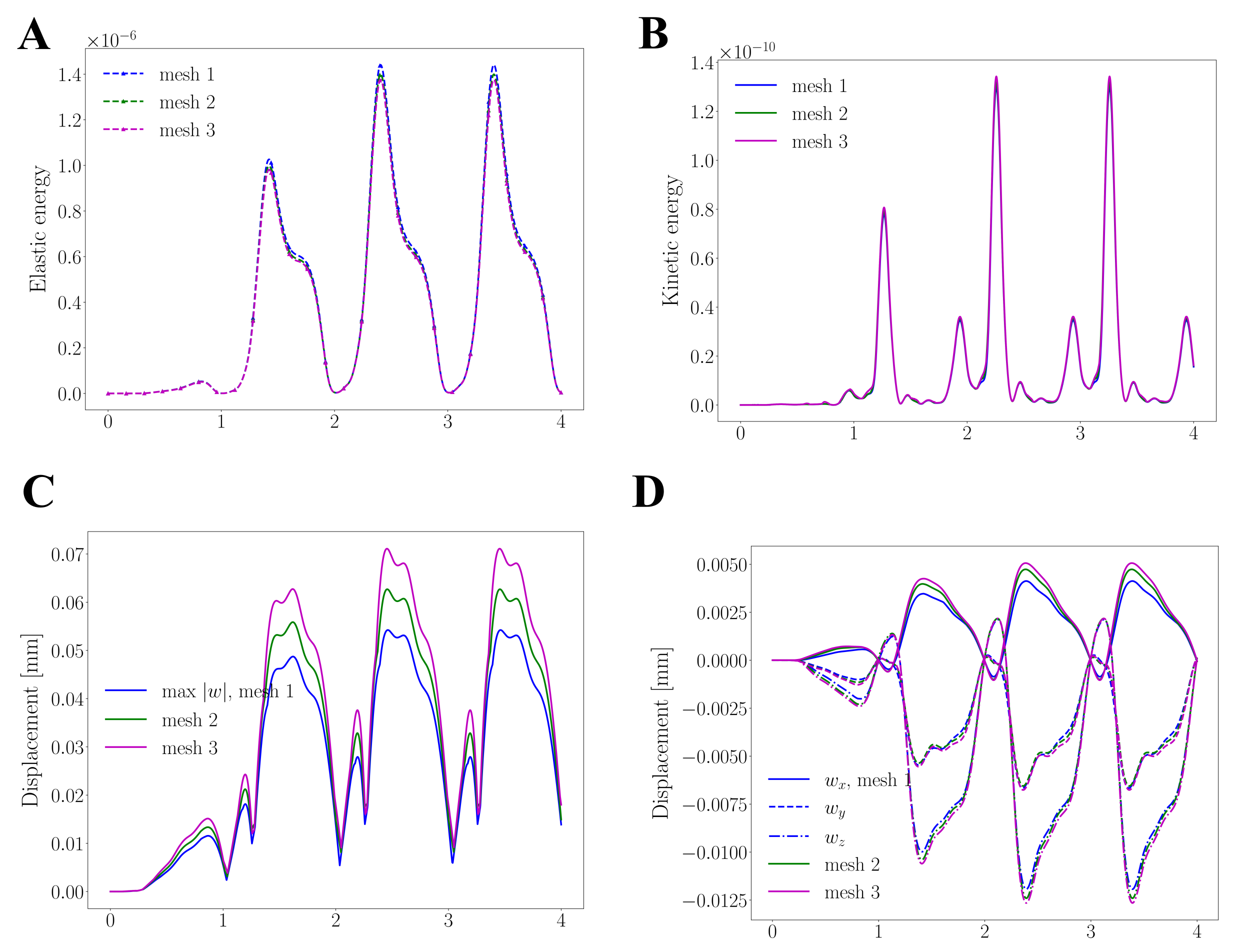}
    \caption{Convergence under mesh refinement for the linear elastodynamics model
    using quadratic polynomial basis functions ($p=2$). The standard mesh is mesh 1.}
    \label{fig:verification_deformation_mesh}
\end{figure}

\end{document}